\documentclass[twoside,onecolumn]{article}

\usepackage{blindtext} 

\usepackage[sc]{mathpazo} 
\usepackage[T1]{fontenc} 
\usepackage{microtype} 

\usepackage[english]{babel} 

\usepackage[hmarginratio=1:1,top=22mm,columnsep=20pt]{geometry} 
\usepackage[hang, small,labelfont=bf,up,textfont=it,up]{caption} 
\usepackage{booktabs} 

\usepackage{lettrine} 

\usepackage{enumitem} 
\setlist[itemize]{noitemsep} 

\usepackage{abstract} 

\usepackage{titlesec} 
\renewcommand\thesection{\Roman{section}} 
\renewcommand\thesubsection{\roman{subsection}} 
\titleformat{\section}[block]{\large\scshape\centering}{\thesection.}{1em}{} 
\titleformat{\subsection}[block]{\large}{\thesubsection.}{1em}{} 

\usepackage{fancyhdr} 
\usepackage{titling} 

\usepackage{hyperref} 

\usepackage{natbib}
\usepackage{graphicx}
\usepackage{amsmath}

\pretitle{\begin{center}\Huge\bfseries} 
\posttitle{\end{center}} 
\title{Constraining stochastic parametrisation schemes using high-resolution simulations} 
\author{%
\textsc{H.~M.~Christensen}\thanks{A thank you or further information} \\[1ex] 
\normalsize Atmospheric, Oceanic and Planetary Physics, University of Oxford, UK. \\ 
\normalsize \href{mailto:hannah.christensen@physics.ox.ac.uk}{hannah.christensen@physics.ox.ac.uk} 
}
\date{\today} 
\renewcommand{\maketitlehookd}{%
\begin{abstract}
\noindent 
Stochastic parametrisations are used in weather and climate models to improve the representation of unpredictable unresolved processes. When compared to a deterministic model, a stochastic model represents `model uncertainty', i.e., sources of error in the forecast due to the limitations of the forecast model. We present a technique for systematically deriving new stochastic parametrisations or for constraining existing stochastic approaches. A high-resolution model simulation is coarse-grained to the desired forecast model resolution. This provides the initial conditions and forcing data needed to drive a Single Column Model (SCM). By comparing the SCM parametrised tendencies with the evolution of the high resolution model, we can estimate the error in the SCM tendencies that a stochastic parametrisation seeks to represent. We use this approach to assess the physical basis of the widely used Stochastically Perturbed Parametrisation Tendencies (SPPT) scheme. We find justification for the multiplicative nature of SPPT, and for the use of spatio-temporally correlated stochastic perturbations. We find evidence that the stochastic perturbation should be positively skewed, indicating that occasional large-magnitude positive perturbations are physically realistic. However other key assumptions of SPPT are less well justified, including coherency of the stochastic perturbations with height, coherency of the perturbations for different physical parametrisation schemes, and coherency for different prognostic variables. Relaxing these SPPT assumptions allows for an error model that explains a larger fractional variance than traditional SPPT. In particular, we suggest that independently perturbing the tendencies associated with different parametrisation schemes is justifiable, and would improve the realism of the SPPT approach.

\end{abstract}
}

\begin{document}

\maketitle



\section{Introduction}


Weather and climate projections, spanning timescales from a few days to many decades, are routinely presented in a probabilistic manner \citep{houghton1990,buizza2017}. Such probabilistic predictions are required to support decision-making. This enables preventative action to mitigate the impacts of extreme weather events, or future climate change. However, overconfident predictions can lead users to make decisions that are costly, while the benefits of those decisions are never realised \citep{murphy1977}. To be useful, a forecast must be reliable --- it must accurately represent the likelihood of the forecast event --- and so must represent all sources of uncertainty in the forecast \citep{stensrud2000}. 

A major source of error in both weather and climate prediction are the approximations made when developing the forecast model \citep{hawkins2009}. In particular, limited computer resources lead to the simplified representation of unresolved small-scale processes, such as convective clouds and turbulent transport, through parametrisation schemes. There is much debate as to the optimal representation of this \emph{model uncertainty}, and several methods have been proposed \citep{stainforth2005,bowler2008,rougier2009,kirtman2014,ollinaho2017,leutbecher2017}. An attractive solution, initially proposed for use in weather forecasts, is the use of stochastic parametrisations. Here, atmospheric processes are represented as a combination of a predictable deterministic and an unpredictable stochastic component. Stochastic parametrisations have revolutionised probabilistic weather prediction \citep{palmer2009}. They are now ubiquitous in operational forecasting centres worldwide \citep[e.g.][]{sanchez2016}, and have also been widely adopted by seasonal forecasting systems \citep[e.g.][]{charron2010}. They have been shown to outperform other representations of model uncertainty on weather and seasonal timescales \citep{christensen2015d,weisheimer2011}. Recent work has considered the impact of stochastic parametrisations in climate models, where they have been found to substantially improve both mean state and variability \citep{wang2016,christensen2017,berner2017,davini2017,strommen2018}.

\begin{figure}
  \centering
  \includegraphics[width=0.48\textwidth]{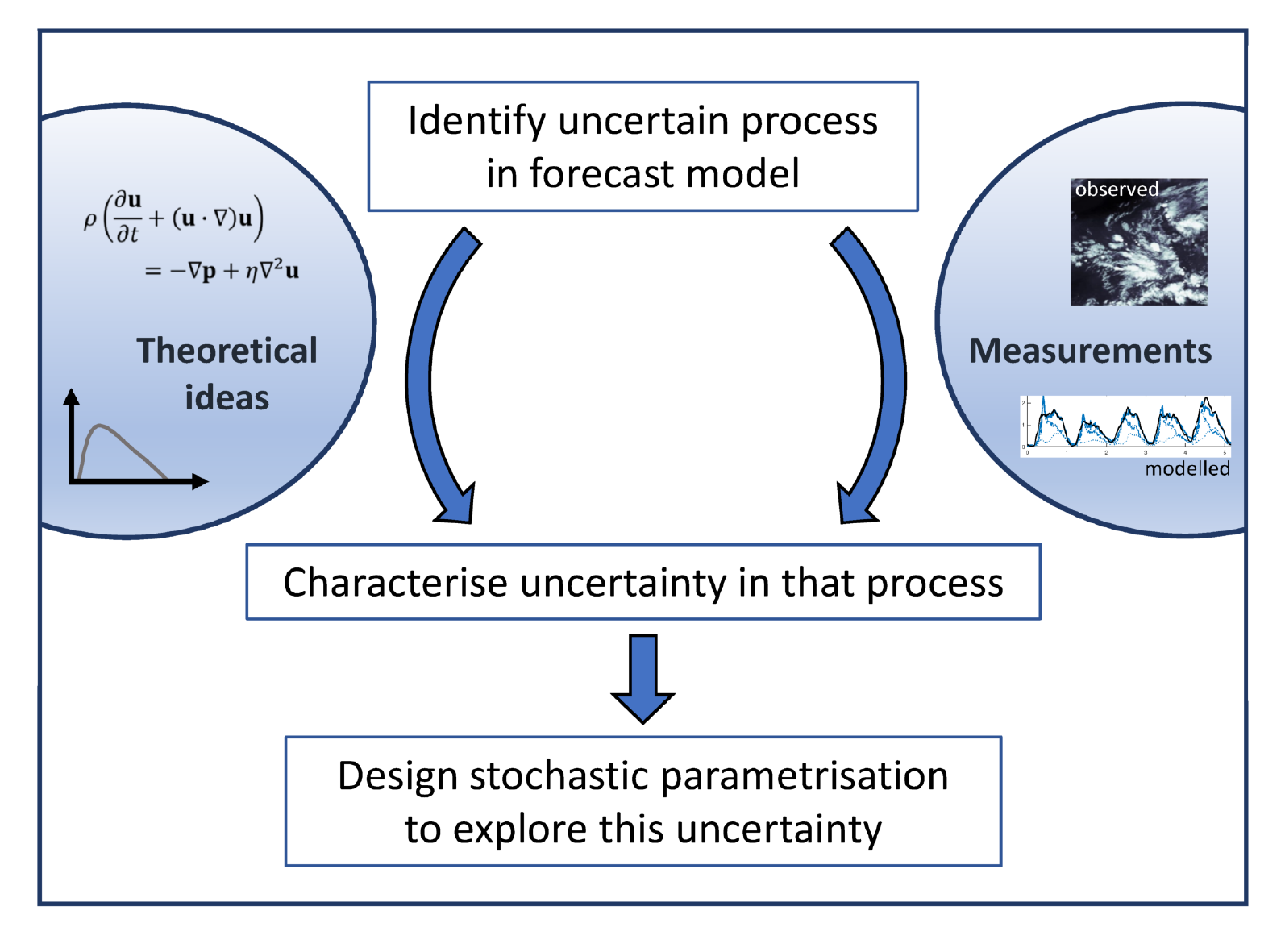}
    \caption{Schematic illustrating the process of developing stochastic parametrisation schemes.}
  \label{fig:schematic}
\end{figure}

As summarised in Figure~\ref{fig:schematic}, the starting point for all stochastic schemes should be identifying a source of uncertainty in models, be this error in the representation of a specific process such as sub grid-scale turbulent mixing \citep{suselj2014} or variability in air-sea fluxes \citep{bessac2019}, or a collection of processes, such as uncertainty in the net parametrised physics tendency \citep{buizza1999}. Having identified the model error that leads to uncertainty in the forecast, the characteristics of that model error must be predicted through theory, or assessed through measurements. This allows for a statistical representation of that error to be included into the forecast model. 

Several schemes have been proposed whereby the characteristics of model error are predicted through theoretical understanding. For example, \citet{plant2008} propose a stochastic convection parametrisation based on the theory of \citet{craig2006}. \citet{khouider2010} propose a stochastic multicloud model based on an understanding of tropical convection, with the transition rates between cloud types set by rules of thumb. \citet{ollinaho2017} propose to stochastically vary 20 parameters in the European Centre for Medium-range Weather Forecasts (ECMWF) model, where the parameters were identified by experts as being sources of uncertainty and the optimal magnitude of perturbation was tuned to maximise forecast skill. However, even with a physical foundation, these approaches tend to contain one of more parameters that must be estimated or tuned. To fully characterise model error and constrain these variables, we can augment such theoretical ideas with measurements. For example, \citet{ollinaho2013b} describe the use of a Bayesian parameter estimation framework to quantify the uncertainty in four parameters within the convection parametrisation scheme, which was used by \citet{christensen2015d} to develop a well-constrained stochastically perturbed parameter scheme. \citet{dorrestijn2015} use observational data to estimate the transition probabilities between cloud types to underpin the stochastic multicloud approach of \citet{khouider2010}. 

Recent years have seen a large increase in the production of very high resolution atmospheric simulations. Continued increases in computational power have allowed for increases in domain size and duration of simulations, with resolutions regularly reaching convection permitting, if not convection resolving, scales \citep{holloway2012,satoh2014,schalkwijk2015,heinze2017,satoh2017}. The availability of such datasets opens up the option of using high-resolution simulations as a proxy for the `true atmosphere', and identifying the difference between a low resolution forecast model and a high-resolution simulation as the model error that a stochastic parametrisation seeks to represent \citep{shutts2007,shutts2014}. This allows for the derivation of data-driven stochastic representations of this sub-grid variability \citep{dorrestijn2013,portamana2014,cooper2015,bessac2019}. To date, this approach has been used to construct stochastic parametrisation schemes that are independent from (i.e. an alternative to) existing deterministic parametrisations.

In this study we revisit the use of very high resolution datasets to evaluate and constrain stochastic parametrisation schemes. Instead of deriving an independent stochastic scheme, we seek to characterise the error in existing deterministic parametrisations. We then assess whether the deterministic schemes could be augmented by a stochastic term to represent uncertainty in the parametrised tendencies. We follow the technique proposed in \citet{christensen2018b}, whereby existing high-resolution simulations are coarse-grained to provide the input for a low-resolution single column model (SCM), which provides the parametrised tendencies. As a case study, we use this coarse graining methodology to assess the widely used `Stochastically Perturbed Parametrisation Tendencies' (SPPT) scheme. While the SPPT scheme was proposed based on theoretical ideas, the foundational assumptions of the scheme have never been assessed. The coarse-graining methodology will probe the validity of this underlying theory.

We present the SPPT scheme in Section~\ref{sec:SPPT}. In section~\ref{sec:cg} we review the coarse-graining methodology presented in \citet{christensen2018b}, and highlight adaptations to the methodology necessary for this study. In section~\ref{sec:assessSPPT} we begin to assess the main assumptions underlying SPPT, and highlight successes of the scheme. In section~\ref{sec:beyondSPPT} we demonstrate that other assumptions made in SPPT are not justifiable. In section~\ref{sec:discussion} we discuss the results, and conclude in section~\ref{sec:concs} by recommending simple changes to the SPPT approach.


\section{Chosen application: the Stochastically Perturbed Parametrisation Tendencies Scheme} \label{sec:SPPT}


The stochastically perturbed parametrisation tendencies (SPPT) scheme is an attractive approach due to its ease of use and beneficial impact on ensemble forecast reliability \citep{palmer2009,weisheimer2014a}. It runs in conjunction with operational physics parametrisation schemes, so can be easily adapted for use in different models. It is widely used in weather and seasonal forecasting centres worldwide, including at the European Centre for Medium-range Weather Forecasts (ECMWF) \citep{buizza1999,palmer2009}, the U.K. Met Office \citep{sanchez2016}, the Japan Meteorological Agency \citep{yonehara2011}, in the Application of Research to Operations at Mesoscale (AROME) model \citep{bouttier2012} and the Weather Research and Forecasting (WRF) model \citep{berner2015}, and in the Community Earth System Model \citep{christensen2017} and EC-Earth \citep{davini2017} climate models. While the essence of the scheme is the same across models, there are differences in precise implementation. The following discussion presents the details of the scheme as implemented at ECMWF \citep{palmer2009,leutbecher2017}.

The SPPT scheme addresses model uncertainty due to the parametrisation process. It does this by perturbing the sum of the parametrised physics tendencies using multiplicative noise:
\begin{align}
\mathbf{T}_X &:= \frac{\partial X}{\partial t} \nonumber \\ 
           &= \mathbf{D}_X + (1+e)\sum_{i=1}^I {\mathbf{P}_{i,X}} \nonumber \\ 
           &= \mathbf{D}_X + (1+e)\mathbf{P}_X. \label{eq:SPPT}
\end{align}
where $\mathbf{T}_X$ is the total vector tendency in $X$, as a function of model level at a particular spatial grid point. $\mathbf{D}_X$ is the vector tendency from the dynamics, $\mathbf{P}_{i,X}$ is the vector tendency from the $i\mathrm{th}$ physics scheme, and $e$ is a zero mean random perturbation. There are $I=5$ key physics schemes in the IFS: radiation (RDTN); turbulence and orographic gravity wave drag (TGWD); non-orographic gravity wave drag (NOGW); convection (CONV); large-scale water processes (LSWP). The perturbation $e$ is constant in the vertical though it is tapered in the boundary layer and stratosphere. The tapering in the boundary layer is to avoid exciting numerical instabilities in the model, while in the stratosphere, uncertainty in the parametrised tendencies is believed to be small.

The SPPT scheme perturbs the tendency for four prognostic variables: $X$ = temperature ($T$), zonal and meridional wind speed ($U$ and $V$ respectively), and humidity ($q$). Each variable tendency is perturbed using the same random number field. The perturbation field is generated using a spectral pattern generator. The pattern at each time step is the sum of three independent random fields with horizontal correlation scales of 500, 1000 and 2000 km. These fields are evolved in time using an AR(1) process with time scales of 6 hours, 3 days and 30 days respectively. The fields have standard deviations of 0.52, 0.18 and 0.06 respectively. It is expected that the smallest scale (500 km and 6 hours) will dominate at a 10 day lead time, while the larger scale perturbations are important for monthly and seasonal forecasts.

Underpinning SPPT are several theoretical statements \citep{buizza1999}:
\begin{enumerate}
\item The larger the parametrised tendencies, the larger the potential random error. This is based on the concept that random model error arises due to unresolved organisation of sub-grid processes. A higher degree of sub-grid organisation will increase the mean sub-grid tendency, and also increase the variability.
\item Random error due to the parametrisation process will be coherent between the different physics parametrisation schemes. This is such that the balance between tendencies associated with different physics schemes is retained.
\item Random error due to the parametrisation process will be coherent between the parametrised tendencies for different prognostic variables ($T$, $U$, $V$, $q$). This ensures physical consistency in the model.
\item The random error is coherent across large spatial and temporal scales, due to the source of the random error being the lack of sub-grid organisation in the forecast model. Furthermore, the truncation of the model equations of motion is expected to introduce errors on both larger and smaller scales than the truncation scale, introducing correlations into the random error.
\end{enumerate}

Some evidence supporting the first statement has been provided by past coarse graining studies. \citet{shutts2007} defined an idealised cloud resolving model (CRM) simulation as truth. The high-resolution fields and their tendencies were coarse grained to the resolution of a NWP model to study the sub-grid scale variability which a stochastic parametrisation seeks to represent. The `true' convective heating on the coarse grid was calculated by averaging the convective heating over nine fine grid boxes. This was compared to the heating calculated from a convection parametrisation scheme on the coarse grid. The validity of the multiplicative noise in the SPPT scheme was analysed by studying histograms of the coarse grained `true' heating conditioned on different ranges of the parametrised heating. The mean and standard deviation of the true heating are observed to increase as a function of the parametrised heating, providing some support for the SPPT scheme. However, \citet{shutts2007} do not quantify whether the relationship is linear, as assumed by SPPT. Furthermore, \citet{shutts2007} only consider a single level in the atmosphere, and so do not test the coherency of the error in the vertical.

Despite some evidence in support of the first statement, coarse graining studies have indicated that the second underpinning statement may be less valid. A different coarse graining study by \citet{shutts2014} estimated the standard deviation of the error for each physics tendency as a function of the parametrised tendency. The tendencies from the IFS at T1279 (16km) were defined to be ``truth'', and were compared to forecast tendencies from the IFS at T159 (130km). The study revealed that the different schemes have different error characteristics, with the uncertainty in the cloud and convection tendencies being much larger than the radiation tendency. \citet{shutts2014} also found that the standard deviations of the cloud and convection tendencies were a non-linear function of the parametrised tendency \citep{shutts2014}. However, the T1279 simulation used as `truth' is relatively low resolution, and includes parametrised convection. It is not clear how this would impact the analysis. 

The second statement above also assumes that the errors from each physics parametrisation scheme are perfectly correlated --- one random number field is used to perturb all schemes. This has not been assessed. It is unlikely that uncertainties in the different processes are precisely correlated, as modelled by SPPT. An alternative is to use independent random fields for each physics tendency. This ``independent SPPT'' (iSPPT) was considered by \citet{christensen2017b}, and resulted in a significant improvement in the reliability of ensemble forecasts, particularly in regions with significant convective activity.

Regarding the third statement, the original implementation of SPPT perturbed different prognostic variable tendencies with different random numbers \citep{buizza1999}. The move to using a single pattern to perturb all prognostic variable tendencies was proposed later to ensure physical consistency \citep{palmer2009}, for example, accounting for the relationship between temperature and humidity tendencies for thermodynamic processes. Moving to a single pattern was found to benefit (reduce) the frequency of strong precipitation events in the forecast \citep{palmer2009}. Nevertheless, the validity of the third statement has not been tested.

The validity of the fourth statement has also not been tested. It is well known that stochasticity must be applied on large spatial and temporal scales to noticeably impact the forecast, as grid-scale noise is readily dissipated by the model equations. However, it is not clear whether there is a physical origin for these large spatio-temporal correlation scales. The chosen scales in SPPT have been simply tuned to give the best results.


\section{The Coarse-Graining Framework} \label{sec:cg}


In this study, we use a high-resolution convection permitting atmospheric simulation to address the theoretical ideas underpinning SPPT. We consider the multiplicative nature of SPPT, the coherency of the uncertainty arising from different physics schemes, the coherency of the uncertainty in different prognostic tendencies, and the existence of large spatio-temporal correlation scales in the optimal perturbation.

The high resolution simulation used here was one of several simulations produced by the UK `Cascade' project, funded by the Natural Environment Research Council. The chosen simulation was produced using the UK Met Office Unified Model (MetUM) at 4 km resolution, covering the Indo-Pacific Warm Pool region, 20$^\mathrm{o}$S--20$^\mathrm{o}$N, 42--177$^\mathrm{o}$E \citep{holloway2012}. The model is semi-Lagrangian and non-hydrostatic, and uses Smagorinsky sub-grid mixing in the horizontal and vertical dimensions. At 4~km resolution, the model is `convection permitting'. The closure of the convection scheme is adapted such that almost all rainfall is generated explicitly, with the convection scheme only active in weakly unstable cases\footnote{Less than $0.1\%$ of precipitation in the Cascade simulation is due to parametrised convection.}. The model has 70 terrain following levels in the vertical with a model top at 40 km, and uses a time step of 30s. The lateral boundary conditions were provided by relaxing the simulation to a 12 km parametrised run through a nudged rim of 8 grid points. The simulation begins on 6 April 2009, and spans 10 days. The start date was selected to study an active Madden-Julian oscillation (MJO) event. The data is stored at full resolution in space, and once an hour in time, and is available on request from the NERC Centre for Environmental Data Analysis (CEDA). For further details of the simulation, see \citet{holloway2012}.

The high-resolution simulation realistically simulates tropical meteorology, as reported on in \citet{holloway2012,holloway2013,holloway2015}. The convection permitting simulation showed substantial improvements over simulations with parametrised convection, including simulating a realistic rainfall distribution \citep{holloway2012}, vertical heating structure \citep{holloway2015}, relationship between precipitation rate and tropospheric humidity  \citep{holloway2012,holloway2013}, and realistic generation of eddy available potential energy  \citep{holloway2013}. The ability to simulate these basic physical relationships allows for a realistic MJO, including simulating a degree of convective organisation, MJO strength, and propagation speed that match observations \citep{holloway2013}. However, the simulation does not conserve moisture, with the advection scheme creating spurious rainfall \citep{holloway2015}. Fortunately, the excess rainfall is not associated with an anomalous heating, so a sufficient solution is to scale back the Cascade rainfall when comparing to observations. The simulation captures dynamical features, including a realistic representation of horizontal and vertical wind speeds, though ascent is more spatially confined in observations than in the simulation \citep{holloway2013}. The first day of the Cascade simulation showed a very strong spin-up \citep{holloway2012}. We therefore discard the first day, and focus our analysis on the remaining nine days.

This study will treat this existing high-resolution simulation as the `truth' which we would like a low resolution forecast model to be able to mimic. We first coarse-grain the high resolution simulation to the resolution of the forecast model. Ideally, the low resolution forecast model would be able to predict these coarse-grained fields. The difference between the low resolution forecast model and this coarse-grained `truth' is defined to be the model error in the low resolution forecast model. It is this error that a stochastic parametrisation seeks to represent. By characterising the statistics of the model error, we can design a stochastic parametrisation scheme to represent this model error in forecasts, following Figure~\ref{fig:schematic}.

\subsection{The IFS SCM}

This study will evaluate the model error of the ECMWF model, the Integrated Forecasting System (IFS), model version CY40R1, at $\mathrm{T_L}$639 resolution (approximately 30~km grid box) with 91 vertical levels and a timestep of 15 minutes. This is a typical resolution used in a global ensemble prediction system. This version of the IFS SCM has been released through the OpenIFS project. To combine the coarse-graining procedure with the low-resolution forecast model, we adapt the methodology described in \citet{christensen2018b}. Instead of considering forecasts made with the global IFS, the IFS Single Column Model (SCM) is used to integrate forward the equations of motion in each coarse-scale grid column.

The IFS SCM represents a single column taken from the global IFS model. The code base is the same as for the global IFS model, and includes the atmospheric physics parametrisations and the land surface scheme. It contains a simplified set of dynamical equations, and requires specification of external dynamical forcing fields including the vertical velocity, geostrophic winds, and advective tendencies of $T$, $U$, $V$, and $q$. In addition, the IFS SCM requires initial conditions for the atmospheric column, and boundary conditions describing the sea surface temperature (SST), orography, vegetation, and surface fluxes. Given these input fields, the SCM predicts the future evolution of the atmospheric column. This includes a decomposition of the change in the prognostic variables ($T$, $U$, $V$, $q$) into a component from each physical parametrisation scheme and a component from the dynamics (i.e. advection and diffusion).

\subsection{Coarse-graining the Cascade dataset}

The coarse-graining methodology is detailed in \citet{christensen2018b}. An overview is provided here for ease of reference.

The IFS $\mathrm{T_L}$639 reduced gaussian grid is used to define the latitude and longitude co-ordinates that make up the coarse-scale SCM grid.  The fields from Cascade are coarsened onto the $\mathrm{T_L}$639 grid using local area averaging. This allows for high-resolution grid boxes to contribute a fractional component to several coarse-resolution grid boxes:
\begin{equation}
\overline{\psi}_{n,k} = \sum_f W_{n,f} \psi_{f,k} \label{eq:av}
\end{equation}
where $\psi_{f}$ denotes the field on the fine grid and $\overline{\psi}_n$ denotes the field on the coarsened grid. The coarse (fine) grid box is identified by the index $n$ ($f$). $W_{n,f}$ indicates the fraction of fine grid box $f$ within coarse grid box $n$, and the vertical level of the field is indicated by index $k$.

Both the fine- and coarse-resolution datasets are defined on model levels, and interpolation must also be performed in the vertical. We choose to perform vertical interpolation second, after first averaging horizontally across each model level. The first field to be coarsened in this way is the surface pressure. The low-resolution surface pressure field is combined with the ECMWF hybrid height coefficients, $A_k$ and $B_k$, to define the pressure on the ECMWF hybrid model levels. The coarse-grained Cascade dataset is then interpolated, logarithmically in pressure, from the Cascade model levels to the ECMWF model levels. The Cascade model top is at 40~km altitude. Above this level, the fields are padded using ECMWF operational analysis data. The Cascade and analysis datasets are smoothly blended over five levels. Finally, a 9-point gaussian smoother is applied to all initial condition fields after coarse graining. This removes small scale features present in the Cascade simulation that are unresolved on the low resolution grid, and which therefore appear as grid-point noise.

The IFS SCM assumes all dynamical forcing fields are instantaneous. The advected tendencies of the prognostic variables ($T$, $U$, $V$, $q$) are calculated along IFS model levels from the coarsened fields:
\begin{equation}
\mathrm{adv}(\psi)|_{n,k} = - \overline{\mathbf{u}}_{n,k}\cdot\overline{\nabla}_k(\overline{\psi_{n,k}})
\end{equation}
for variable $\psi$. A centred finite difference scheme is used to estimate the vector gradient in $\psi$ before the dot product is taken with the coarse-grained vector wind field, $\overline{\mathbf{u}}_{n,k}$. The required geostrophic wind forcing and vertical velocity forcing are also evaluated using the coarse-grained fields: see \citet{christensen2018b} for more details.

The constant boundary fields required by the SCM are taken from the ECMWF archive at $\mathrm{T_L}$639 resolution, ensuring the SCM has the same boundary conditions as the global model. Interactive land surface processes are turned off in the SCM, and replaced with time varying latent and sensible heat fluxes from the Cascade simulation.

The coarse-grained Cascade data sets required to drive the IFS SCM are archived at the Natural Environmental Research Council Centre for Environmental Data Analysis \citep{christensen2018a}.

\subsection{Experimental details}

An IFS SCM integration is initialised once an hour for each grid box of the coarse-grained Cascade simulation, starting at 00 UTC on 7 April 2009. Each simulation lasts two hours, or eight SCM timesteps. The IFS shows a spin-up period over the first few timesteps, with anomalously high rain rates and associated moisture and temperature tendencies \citep{christensen2018b}. The first hour of each SCM simulation is therefore discarded, and the second hour considered for analysis. This is to focus on error statistics relevant to the bulk of the model simulation. It is necessary to consider the cumulative error over four timesteps because the Cascade fields are saved once an hour. The SCM is not nudged to the observed Cascade fields, but instead evolves freely. The lowest 60 IFS model levels are considered, excluding those above the Cascade model top. Note that model level 1 corresponds to the IFS model top, while model level 91 is closest to the ground. For a conversion between model levels and characteristic pressure levels, see Table S1 in the online supporting information.

The high resolution Cascade simulation is nudged towards a lower resolution simulation over a rim of points 32km wide. We therefore discard a rim of two coarse-grained points before analysis. The remaining domain spans 138 latitudinal by 476 longitudinal coarse-grained points, i.e. in excess of 65,000 independent SCM simulations per time step. 


\section{Assessing SPPT: Multiplicative Noise} \label{sec:assessSPPT}


\subsection{Testing the multiplicative noise hypothesis} \label{sec:mult}

\begin{figure*}
  \centering
  \includegraphics[trim=0cm 0cm 0cm 0cm, clip=true, width=0.98\textwidth]{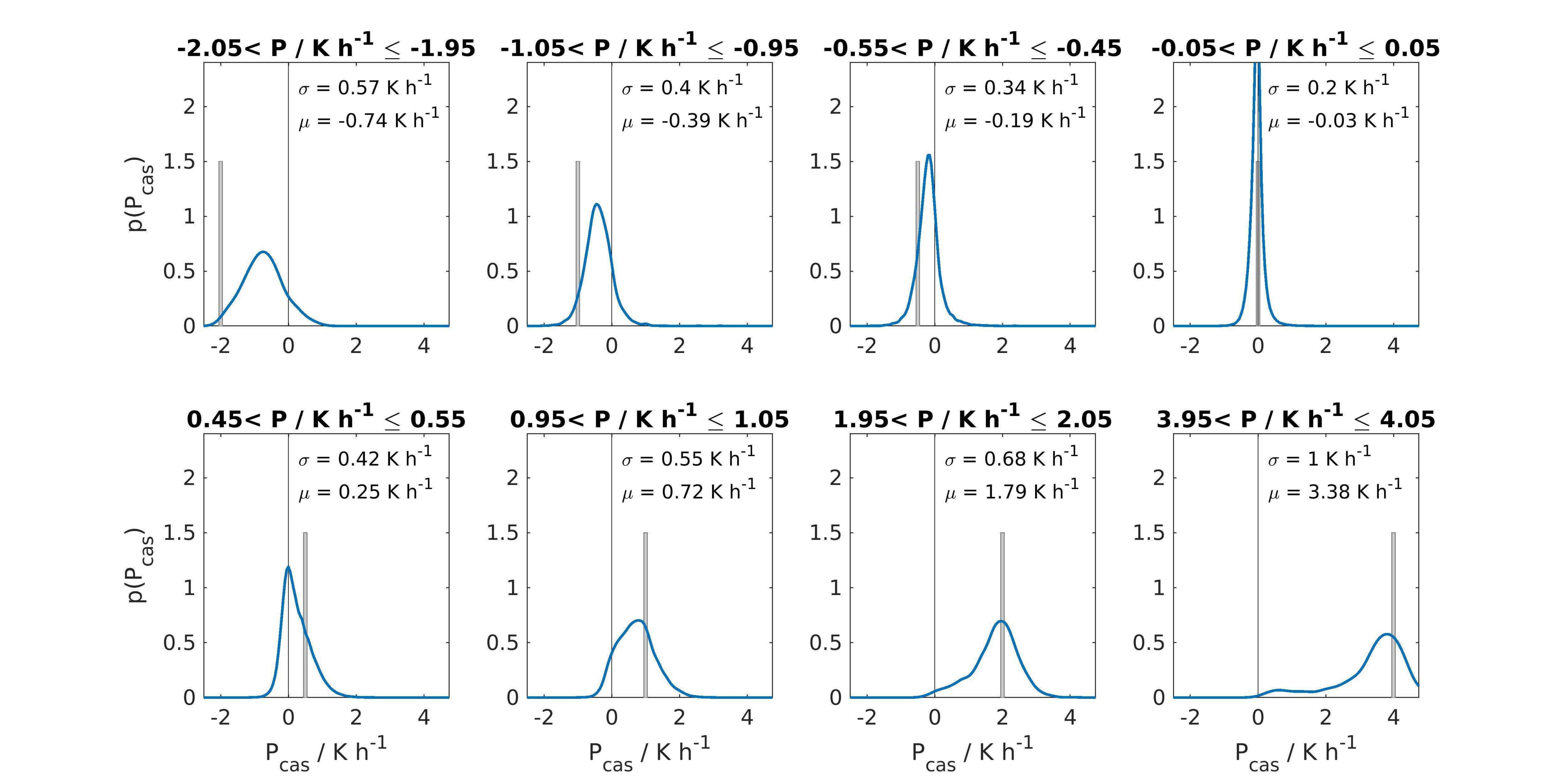}
    \caption{Testing the multiplicative noise hypothesis. Results are shown for the $T$ tendency at model level 77 (approximately 850 hPa). Each subplot shows the distribution of $\mathbf{P}_{cas}$ conditioned on the physics tendency predicted by the SCM indicated by the title of each panel. The mean and standard deviation of the conditional $\mathbf{P}_{cas}$ distribution are shown in each panel. The grey rectangle shows the $\mathbf{P}$ distribution for each panel, though note that the height of this distribution is truncated to 1.5 for clarity: it should extend to 10. The number of data points used to estimate the pdf are 131, 4123, 29438, 1114817, 87759, 27608, 4020, and 162 for each panel respectively.}
  \label{fig:hist}
\end{figure*}

The experimental procedure outlined above allows the estimation of all the key terms in the SPPT equation~\ref{eq:SPPT}. The total tendency, $\mathbf{T}$ for each variable is defined as the change in a prognostic variable between two consecutive coarse-grained Cascade fields, $(t, t + 1 \mathrm{hr})$. This is the `target' which the forecast model should be able to predict. The SCM integration produces a tendency from each parametrised physics scheme, $\mathbf{P}_i$ over the same one-hour window, having been initialised at $t - 1  \mathrm{hr}$ and the first hour discarded. The SCM combines the provided forcing files from Cascade to produce a dynamics tendency, $\mathbf{D}$. Each of these tendencies are available for each prognostic variable, ($T$, $U$, $V$, $q$), as a function of model level, across the Cascade domain.

Firstly it is assessed whether multiplicative noise is a suitable model for the uncertainty in the IFS parametrisation schemes. To do so, an observed ``Cascade physics tendency'' is constructed as $\mathbf{P}_{cas} = \mathbf{T} - \mathbf{D}$ \footnote{Note that SPPT assumes there is no error in the dynamics tendency. This assumption is not questioned in this paper.}. Treating each prognostic variable independently, the data are sorted by the predicted SCM physics tendency, and grouped into bins of equal width. Figure~\ref{fig:hist} shows the distributions of $\mathbf{P}_{cas} $ conditioned on $\mathbf{P}$ predicted by the SCM for $T$ tendencies at 850 hPa. The chosen range in $\mathbf{P}$ is indicated by the figure panel titles and by the grey rectangle in each figure. Qualitatively the SCM parametrisation schemes are performing well: the average $\mathbf{P}_{cas}$ is well predicted by the SCM for positive tendencies, though there is a bias in the negative tendencies, with the mean $\mathbf{P}_{cas}$ having a smaller magnitude than $\mathbf{P}$. It is also apparent from figure~\ref{fig:hist} that the uncertainty in the true tendency increases as the tendency increases, as modelled by SPPT. However, the distributions are not gaussian, and there is a non-negligible probability that the `true' tendency has the opposite sign to the predicted tendency. This is not allowed by SPPT. The fourth panel shows the distribution of $\mathbf{P}_{cas}$ when the predicted tendency is zero. The standard deviation is substantial. Multiplicative noise is not able to represent this observed model uncertainty for small tendencies.

\begin{figure*}
  \centering
  \includegraphics[trim=0cm 0cm 0cm 0cm, clip=true, width=0.98\textwidth]{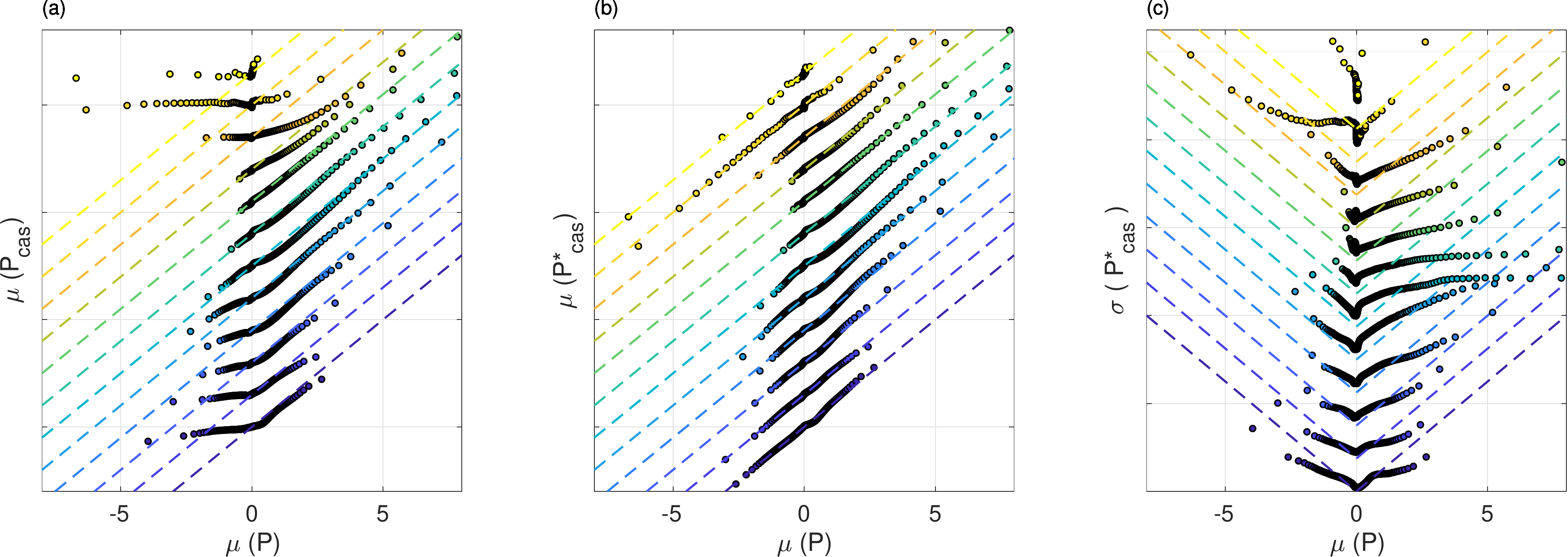}
    \caption{Testing the multiplicative noise hypothesis. Results are shown for the $T$ tendency. (a) The mean of $\mathbf{P}_{cas}$ and (b) the mean of the bias corrected $\mathbf{P^*}_{cas}$, both conditioned on $\mathbf{P}$. The dashed lines indicate the one-to-one line for each level group. (c) The standard deviation of the corrected $\mathbf{P^*}_{cas}$ conditioned on $\mathbf{P}$. The dashed lines have a gradient of 0.5. In all panels the colours indicate the model level group, from dark blue (levels 87--91 / 980--1000 hPa) through to yellow (levels 32--36 / 60--80 hPa). Each level group is plotted with a vertical offset for clarity. The bin edges are defined based on the percentiles of the predicted tendency distribution for each level group, and so are equally populated.}
  \label{fig:hist_manylev}
\end{figure*}

Figure~\ref{fig:hist} only tests the multiplicative noise hypothesis at one height in the atmosphere. Figure~\ref{fig:hist_manylev} summarises this analysis for all model levels for $T$ (the equivalent figures for $q$, $U$, and $V$ are included in the supplementary material, figures S1, S2 and S3). In the vertical, the levels are grouped into 12 groups of five levels, and the data from each group of five levels are binned into 100 equally populated bins. Panel (a) shows the mean of $\mathbf{P}_{cas}$ conditioned on $\mathbf{P}$. If the SCM physics parametrisations are able to predict the `observed' Cascade physics tendency accurately, the scattered points in Figure~\ref{fig:hist_manylev}(a) should lie on the one-to-one dashed line. The positive temperature tendencies are well calibrated across almost the whole vertical domain, while the bias in negative tendencies highlighted in Figure~\ref{fig:hist} is confined to the lower troposphere and stratosphere. 

Stochastic parametrisation schemes are designed to represent random model error. Therefore the systematic biases identified in Figure~\ref{fig:hist_manylev}(a) will be modelled separately. Motivated by figure~\ref{fig:hist_manylev}(a) and adopting the simplest functional form, the systematic difference between the average Cascade physics tendency, $\mu(\mathbf{P}_{cas})$ and the SCM physics tendency $\mathbf{P}$ is modelled as
\begin{align}
\mu(\mathbf{P}_{cas}) &= a_1 \mathbf{P} + m &:= \mathbf{P^*},  \quad \quad \mathbf{P} >0 \\
\mu(\mathbf{P}_{cas}) &= a_2 \mathbf{P} + m &:= \mathbf{P^*},  \quad \quad \mathbf{P} <0
\end{align}
Where a linear functional form has been assumed. This defines the debiased SCM tendency, $\mathbf{P^*}$. The gradient is calculated separately for positive and negative tendencies, but a common intercept ensures a continuous function. The fit parameters are calculated separately for each model level and variable, and are available graphically in the supplementary material, Figure S4. The bias, $\mathbf{b}(\mathbf{P})$, is defined such that $\mu(\mathbf{P_{cas}}) = \mathbf{P^*} = \mathbf{P} - \mathbf{b}(\mathbf{P})$. The SPPT equation~\ref{eq:SPPT} is rewritten to include this representation of both sources of error:
\begin{equation}
	\mathbf{T} = \frac{\partial X}{\partial t} = \mathbf{D} + (1+e)\mathbf{P} - \mathbf{b}(\mathbf{P}) \label{eq:SPPTdb}
\end{equation}
This error model is used for the rest of the paper.

To probe the statistics of the random model error, $e$, we first define a corrected cascade tendency $\mathbf{P^*}_{cas} = \mathbf{P}_{cas} + \mathbf{b}(\mathbf{P})$. Figure~\ref{fig:hist_manylev}(b) shows that this simple linear approach is able to remove the majority of the systematic bias: the tendencies are now well calibrated.

Figure~\ref{fig:hist_manylev} (c) shows the standard deviation of $\mathbf{P^*}_{cas}$ conditioned on $\mathbf{P}$. If multiplicative noise is a good representation of model error in the IFS, the standard deviation should be proportional to the magnitude of the mean tendency, such that the scattered points lie on straight lines through the origin. The gradient of these lines indicates the optimal standard deviation of the stochastic perturbation. Straight lines are plotted in figure~\ref{fig:hist_manylev} (c) to guide the eye. The figure indicates there is some justification for the multiplicative noise hypothesis. The standard deviation is an approximately linear function of the mean SCM tendency for negative tendencies across the troposphere. The relationship is also approximately linear for small positive tendencies up to 750 hPa (the lowest four level blocks). In the mid troposphere, the standard deviation does increase with $\mathbf{P}$, but the relationship is markedly nonlinear. The largest warming tendencies do not show a correspondingly large standard deviation.  In the upper troposphere and lower stratosphere, the linear relationship returns.

\begin{figure*}
  \centering
  \includegraphics[trim=0.0cm 0.0cm 0.0cm 0.0cm, clip=true, width=0.98\textwidth]{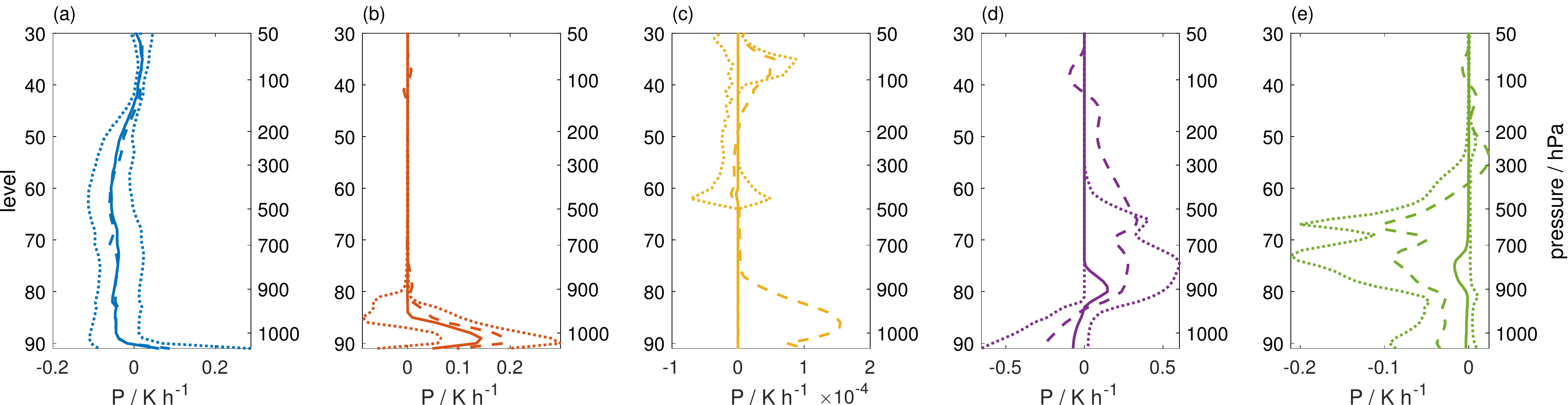}
    \caption{The distribution of tendencies from each parametrisation scheme as a function of model level. (a) Radiation. (b) Turbulence and orographic gravity wave drag. (c) Non-orographic gravity wave drag. (d) Convection. (e) Large scale water processes. In each panel the dotted lines indicate the 16th and 84th percentiles of the distribution as a function of model level; the solid line indicates the median; the dashed line indicates the mean. The mean often differs substantially from the median, indicating a strongly non-gaussian distribution. The right hand vertical axis shows characteristic pressure levels for reference.}
  \label{fig:tends}
\end{figure*}

To attribute these observed characteristics to the behaviour of different parametrisation schemes, we must characterise at which levels each parametrisation is active. Figure~\ref{fig:tends} summarises the probability distribution of T tendencies from each physics scheme as a function of model level. The mean and median of the distributions, and the 16th and 84th percentiles (equal to $\pm 1 \sigma$ for a Normal distribution) are shown. Convection is the dominant parametrisation scheme that warms the mid troposphere, such that the well calibrated positive tendencies at those levels can be attributed to a well calibrated convection scheme. Similarly, the non-linear relationship between standard deviation and mean tendency at these levels is indicative of the statistics of model uncertainty in the convection parametrisation scheme.  The turbulence and gravity wave drag scheme is the primary scheme warming the atmosphere at the lowest model levels. The positive tendencies at these levels appears well calibrated. The uncertainty in the tendencies is a linear function of the mean tendency, though the gradient changes between small tendencies and large tendencies. The TGWD tendency contains contributions from a number of processes including turbulence, mass flux in the boundary layer, and orographic gravity wave drag. It is possible that these different contributions have different uncertainty characteristics.

\subsection{The statistical characteristics of the optimal perturbation}

Section~\ref{sec:mult} indicates that multiplicative noise is a reasonable first-order approximation to the uncertainty in the IFS physics parametrisation schemes, providing support for the use of SPPT. To inform the properties of the stochastic perturbation to be used in SPPT, including magnitude and spatio-temporal correlation scales, we calculate the optimal multiplicative perturbation for every grid point and time step, i.e. the perturbation, $e$, which best maps the forecast tendency onto the `true' tendency. A forecast model cannot predict this optimal perturbation, but the statistics of this perturbation can be included in the forecast model. Each ensemble member in an ensemble forecast will experience a different realisation of this stochastic perturbation. The ensemble will thereby encompass the `true' optimal perturbation in the ensemble, accounting for model uncertainty, and producing a reliable forecast.

To calculate the optimal multiplicative perturbation, we return to the SPPT equation~\ref{eq:SPPT} and rearrange such that the random error in the SCM parametrisation scheme, i.e. the difference between $\mathbf{P}_{cas}$ and $\mathbf{P}$, is represented as a function of the SCM net physics tendency $\mathbf{P}$:
\begin{equation}
	\mathbf{T} - \mathbf{D} -\mathbf{P} +\mathbf{b}(\mathbf{P}) = e \mathbf{P}  \label{eq:SPPTe}
\end{equation}
This is an over-constrained equation for the optimal instantaneous multiplicative perturbation, $e$. Equation~\ref{eq:SPPTe} is solved simultaneously for all prognostic variable tendency vectors. The different prognostic variable tendencies have different units and therefore substantially different magnitudes (e.g. compare Figure~\ref{fig:hist_manylev} with Figures S1, S2 and S3). To remove this dependency on unit and to ensure each variable tendency is weighted equally, each tendency is first divided through by a scale factor $s_X = \sigma(\mathbf{T}_X)$ for variable $X$. In the IFS, SPPT is tapered in the boundary layer and stratosphere for stability reasons. These levels are therefore excluded from the analysis. Since the physics tendencies are smaller at higher model levels (Figure~\ref{fig:tends}) it was found that these higher levels dominated the procedure for fitting $e$. To focus on levels where the tendencies have an appreciable magnitude these levels are also excluded. Only levels between 45 and 87 (inclusive) are used to calculate $e$. The sensitivity of $e$ to the choice of level is shown in the supplementary material, Figure S5. Equation~\ref{eq:SPPTe} is solved by minimising the mean squared residual\footnote{This is achieved using the MatLab backslash operator, which in this case employs a QR decomposition to find the solution.} The solution is the optimal $e$ as a function of horizontal position and time.

\begin{figure*}
  \centering
  \includegraphics[trim=1.0cm 0cm 1.0cm 0cm, clip=true, width=0.98\textwidth]{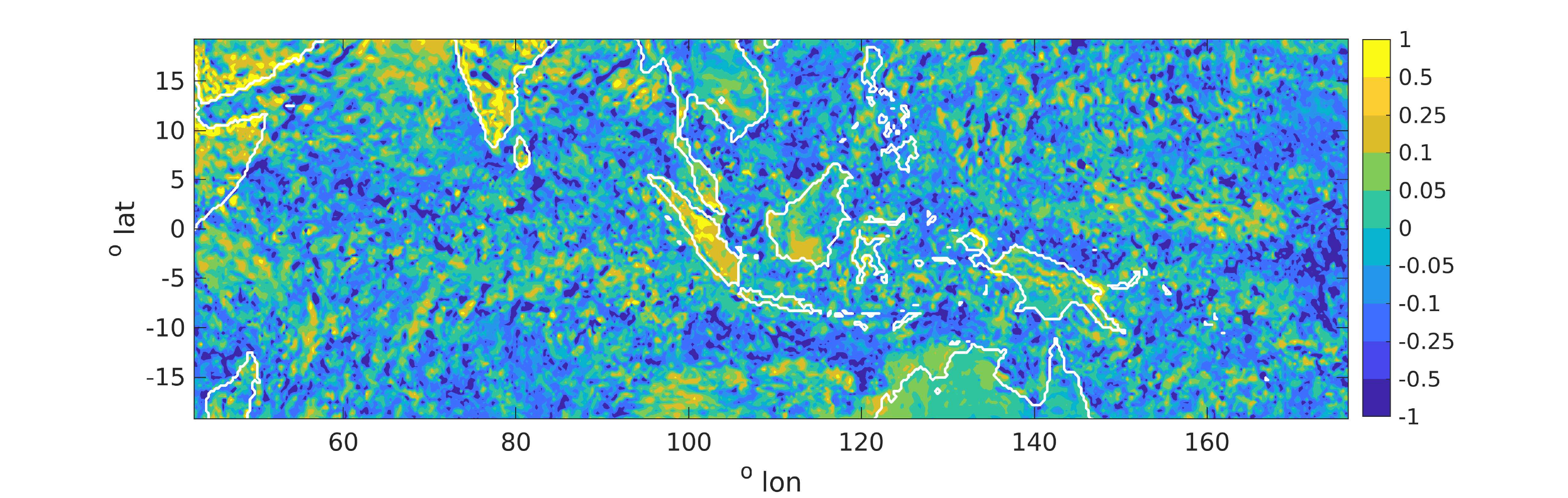}
    \caption{A snapshot of the optimal multiplicative perturbation required to map the SCM forecast onto the observed evolution of Cascade. The SCM forecasts were initialised at 00UTC, 7 April 2009.}
  \label{fig:SPPTe}
\end{figure*}

Figure~\ref{fig:SPPTe} shows a map of the instantaneous optimal multiplicative perturbation, $e$, for forecasts initialised at 00UTC on April 7th 2009. It is evident that there are large-scale correlated structures in $e$. The day-night boundary (at approx. 90 E) is visible in the increased errors over land in the night regions. 

\begin{figure*}
  \centering
  \includegraphics[trim=0.0cm 0.0cm 0.0cm 0.0cm, clip=true, width=0.98\textwidth]{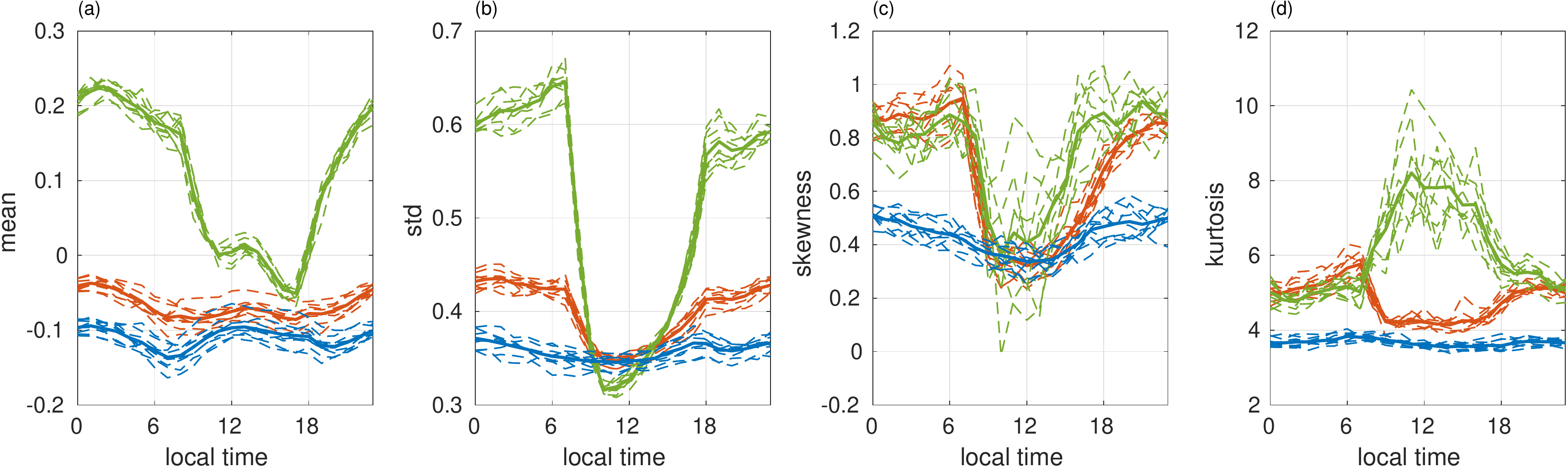}
    \caption{Summary statistics of the distribution of $e$ as a function of local time of day. The statistics are shown calculated over the entire domain (orange) as well as over only land points (green) and over only ocean points (blue). The solid line indicates the average statistics over the entire dataset, while the dashed lines indicate the statistic calculated for each of the nine days separately.}
  \label{fig:SPPTmom}
\end{figure*}

The distribution of $e$ is summarised in Figure~\ref{fig:SPPTmom} in terms of the mean, standard deviation, skewness and kurtosis as a function of local time of day. The dashed lines in Figure~\ref{fig:SPPTmom} show the statistics calculated using each day separately to indicate the variability in each statistic. The statistics aggregated over the entire domain (orange) are compared to those aggregated over only land regions (green) and those aggregated over only ocean regions (blue). Over land regions, the mean perturbation shows a marked diurnal cycle, with perturbations positive on average at night, and zero or slightly negative during the day. The mean perturbation over ocean points is negative, indicating a slight systematic reduction in the magnitude of the physics tendencies of order 10\%. 

The standard deviation over land also shows a marked diurnal cycle, with larger magnitude perturbations at night and smaller magnitude perturbations during the day. In contrast, perturbations over ocean points do not show a strong diurnal cycle. The standard deviation of the perturbation over the ocean is 0.36. Overall, the standard deviation aggregated over the whole domain is 0.40, in contrast to the value of 0.55 used operationally in the IFS. 

The probability distribution of tendencies from each physics scheme, as shown in Figure~\ref{fig:tends}, was recalculated separately for land and ocean points during night and day hours (not shown). While the distribution of parametrised tendencies over ocean does not vary with time of day, the parametrised physics tendencies over land show large differences between day and night. It is not surprising that the characteristics of the error in these processes also varied with the diurnal cycle. At night, the parametrised physics tendencies tend to have smaller magnitude, though still substantial variability. It is possible that a multiplicative representation of error is not able to capture the uncertainty in tendencies with small magnitude, and that an additional, state independent uncertainty could be required.

The calculated $e$ show positive skewness and positive kurtosis over both land and sea regions. This is in contrast to the perturbations used in the IFS, which are normally distributed. The positive skewness indicates that large negative $e$, which could change the sign of the parametrised tendency, are less common than large positive $e$. The kurtosis greater than three indicates the distribution of $e$ has fat tails. Both positive skewness and excess kurtosis indicate the need for large positive perturbations more often than would be modelled by a Gaussian perturbation.

The SPPT scheme used operationally in the IFS uses a sum over three spectral patterns with specified spatial and temporal correlation scales. To inform the optimal length and time scales used in SPPT, $e$ was similarly modelled as a sum over a number of independent patterns.

\begin{figure*}
  \centering
  \includegraphics[width=0.98\textwidth]{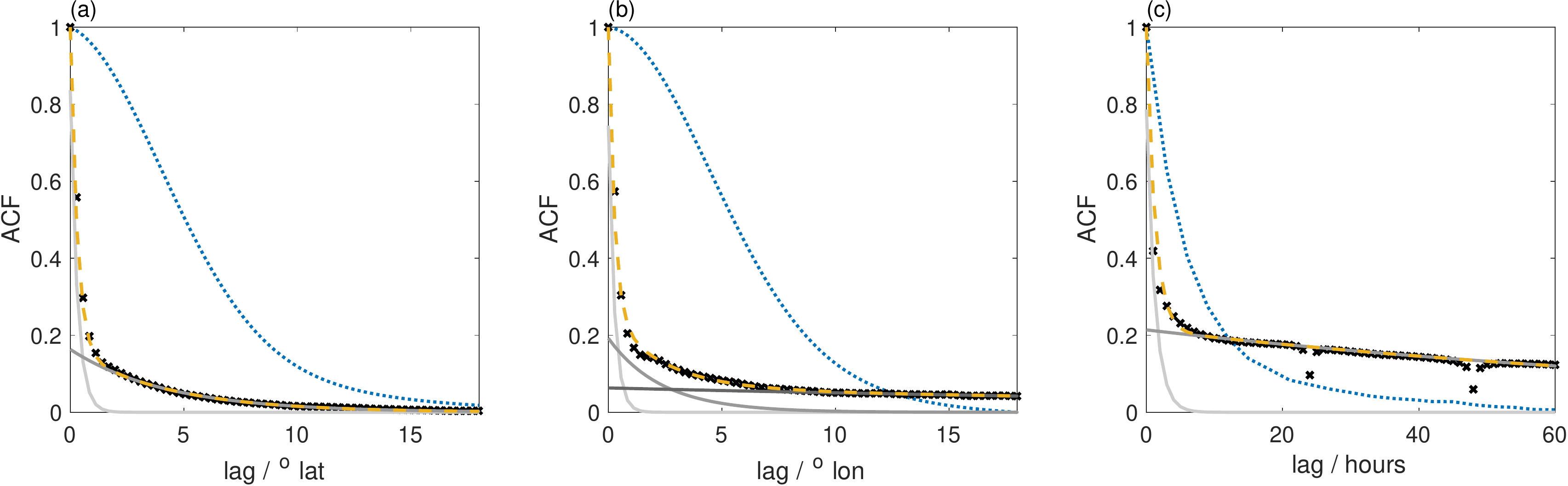}
    \caption{The autocorrelation of $e$ in the (a) latitudinal, (b) longitudinal and (c) temporal dimensions (black crosses). The autocorrelation of $e$ is modelled as the sum over $N$ AR(1) processes, where $N=2$ in the latitudinal and temporal directions and $N=3$ in the longitudinal direction. The individual AR(1) processes are shown as the grey solid lines, whereas the yellow dashed line is the sum over these $N$ processes. The goodness-of-fit is indicated by the match between the yellow dashed line and the black crosses. The autocorrelation of $e$ as used operationally at ECMWF is shown by the blue dotted line.}
  \label{fig:SPPTspattemp}
\end{figure*}

The autocorrelation of $e$ was estimated separately in longitude, latitude, and time. It was found that representing the spatial and temporal autocorrelation functions a sum over $N$ AR(1) processes represented the error processes well. For details of the fitting process, see the Appendix. In longitude, the optimal $N=3$, whereas in latitude and in time, $N=2$. For each case, $N$ was chosen to give the best representation of the estimated correlation function: the domain is large enough in longitude to fit three AR(1) processes, whereas the latitudinal and temporal extents did not permit fitting a third pattern. Figure~\ref{fig:SPPTspattemp} shows the fitted processes for each spatio-temporal direction using data aggregated from over the whole domain. The autocorrelation functions were also estimated separately for land and ocean regions as for Figure~\ref{fig:SPPTmom}. The fitted $e$ over ocean has larger spatial correlation scales than over land, whereas $e$ over land points has a longer decorrelation timescale than over ocean (see supplementary material, Figures S6 and S7).

The parameters fitted in each dimension were combined to produce a spatial and temporal correlation for each scale pattern. Since the longitudinal autocorrelation indicates the existence of a third scale, it is assumed that three patterns exist in each spatio-temporal direction, but the restricted domain size prevents the third pattern from being identified in the latitude and temporal dimensions. The variance ratio for pattern three is therefore taken from the longitudinal fit, and the latitudinal and temporal variance ratios for patterns one and two rescaled to account for the unmeasured third scale. Finally, the relative magnitudes and decorrelation scales of the first two patterns are estimate by averaging the three estimates from each spatio-temporal dimension. For simplicity, this ignores the spatial anisotropy of $e$.

These data are shown in Table ~\ref{tab:SPPTe}, and compared to the values used operationally in the IFS. The first fitted pattern has spatial scale of 32 km and a temporal scale of 1 h. These represent the spatial and temporal resolution of the coarse-scale data, such that this first field corresponds to white noise in time and space. The magnitude of this pattern is smaller than that assigned to the first pattern in operational SPPT. The second pattern accounts for a relatively larger fraction of the variance than in the operational settings, and shows substantial spatial and temporal correlation scales. The fitted spatial scale of 370 km is similar to that used in the first operational SPPT pattern, while the fitted temporal correlation scale of 4.5 d is similar to the second operational SPPT pattern. The third pattern, with expected spatial scales on the order of thousands of km and temporal scales of several weeks, is too large to be constrained by the Cascade dataset, though the third pattern in the longitudinal direction indicates the presence of structures in $e$ with scales of order 8,000 km.

\begin{table}
  \centering
  \begin{tabular}{l | lll | lll }
  \hline
                         &    \multicolumn{3}{l|}{Operational SPPT}   &   \multicolumn{3}{l}{Fitted SPPT}     \\
       \hline
     $\mu(e)$            &  0.0     &         &        &  -0.07    &        &          \\
     $\sigma(e)$         &  0.55    &         &        &   0.40    &        &          \\
     $\sigma_j$          &  0.52,   &  0.18,  &  0.06  &   0.35,   & 0.17,  & 0.10     \\
     $L_j$ (km)          &  500,    &  1000,  &  2000  &   32,     & 370,   &  --      \\
     $\tau_j$            &  6 h,    &  3 d,   &  30 d  &   1.2 h,  & 4.3 d, &  --      \\
  \hline
  \end{tabular}
  \caption{SPPT parameter values for the random fields $j=1, 2, 3$ that comprise the 3-scale pattern used in the IFS: Standard deviation $\sigma_j$, horizontal correlation length $L_j$, time decorrelation scale $\tau_j$. The spatial and temporal scale of the third pattern cannot be accurately estimated due to the limited size of the domain and length of dataset.}
  \label{tab:SPPTe}
\end{table}


\section{Beyond SPPT} \label{sec:beyondSPPT}


Section~\ref{sec:assessSPPT} presented evidence for the multiplicative noise hypothesis, providing some justification for SPPT. Within this framework, the statistical characteristics of the optimal perturbation were estimated. Large scale correlations were revealed in time and space, providing justification for the use of correlated noise in stochastic parametrisation schemes. However it is evident that SPPT is not a perfect representation of uncertainty in IFS. At some vertical levels, the uncertainty in the parametrised tendency is not a simple linear function of the mean tendency. Furthermore, section~\ref{sec:assessSPPT} did not assess all the assumptions made by SPPT, including the coherency of uncertainty in the vertical, between different prognostic variables, and between different parametrisation schemes. These assumptions are tested in the following section.

\subsection{Vertical coherence of perturbations}

We first assess whether a single $e$ is a good representation of model error as a function of model level. In other words, is the vector error in the tendency proportional to the vector tendency. The validity of this assumption was discussed by \citet{leutbecher2017}, who point out that, for example, uncertainty in the shape of a tendency profile cannot be represented using a constant perturbation as a function of height. We probe this question by considering the characteristics of the optimal perturbation fitted separately to each model level, $e_z$. 

\begin{figure*}
  \centering
  \includegraphics[width=0.98\textwidth]{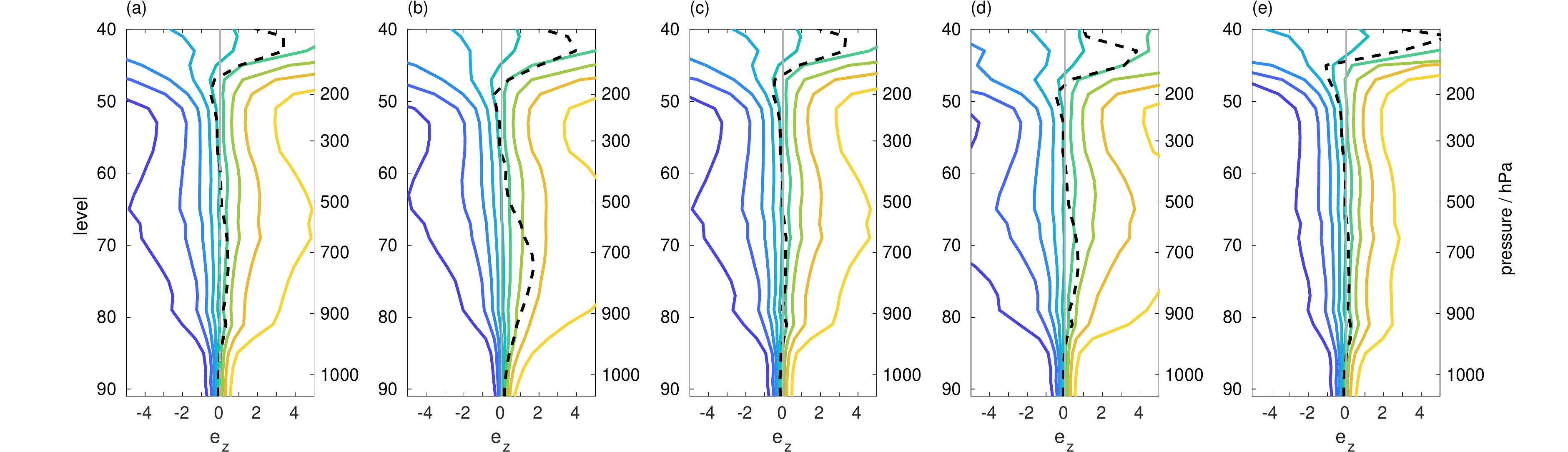}
    \caption{Distributions of optimal $e_z$ as a function of height. The pdf is summarised by showing each decile as a function of model level from dark blue (10th percentile) to yellow (90th percentile). The black dashed line indicates the mean of the distribution. (a) all data, (b) $e_z$ over land points, (c) $e_z$ over ocean points, (d) $e_z$ during the day: 6 am to 6 pm local time and (e) $e_z$ during the night: 6 pm to 6 am local time.  The right hand vertical axis shows characteristic pressure levels for reference.}
  \label{fig:SPPTvert_91_disb}
\end{figure*}

Figure~\ref{fig:SPPTvert_91_disb} summarises the distribution of the optimal $e_z$ as a function of model level in terms of its deciles. It is evident that the characteristics of $e_z$ vary as a function of height in the atmosphere. The standard deviation of $e_z$ is smaller lower in the atmosphere, between levels 85 and 91. The fitted $e_z$ distributions are roughly symmetrical between levels 50 and 91, though for the highest levels in the atmosphere the mean perturbation is positive. Separating the $e_z$ data into (b) land and (c) ocean regions reveals differences. For example, over ocean, $e_z$ for the lowest levels have negative mean and median, which is not the case over land. Over land, the distribution of $e_z$ for levels in the free troposphere is positively skewed, whereas the distribution over ocean is roughly symmetric. Separating the $e_z$ data into (d) day and (e) night reveals smaller uncertainties at night, and higher uncertainties during the day.

\begin{figure*}
  \centering
  \includegraphics[trim=0.05cm 0cm 0cm 0cm, clip=true, width=0.98\textwidth]{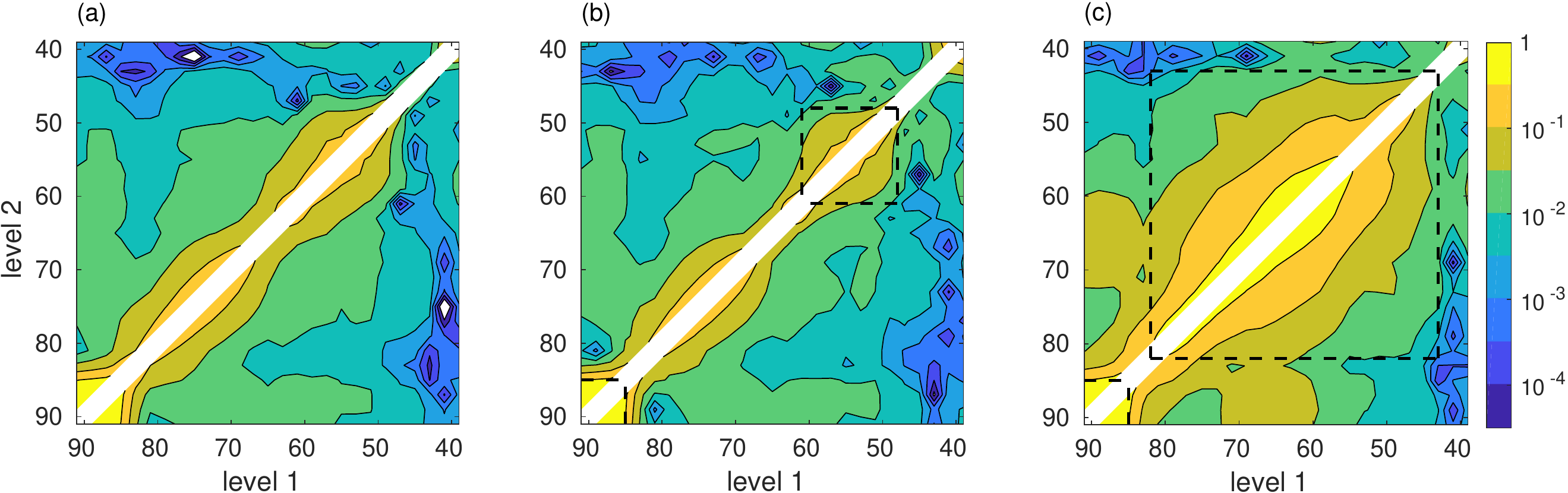}
    \caption{Correlation between the optimal $e_z$ fitted to different model levels. The correlation of $e_z$ with itself at a given level is one, and so has been masked out. (a) all data, (b) $e_z$ during the day: 6 am to 6 pm local time and (c) $e_z$ during the night: 6 pm to 6 am local time. Square features in the correlation indicate a block of levels with high inter-correlations. Dashed black lines highlight such features.}
  \label{fig:SPPTvert_91_corr}
\end{figure*}

Figure~\ref{fig:SPPTvert_91_corr} shows the correlation between perturbations fitted to different levels. If the correlation matrix shows `square' features, this indicates a block of levels that are mutually highly correlated. In general, the perturbations fitted to different levels are weakly correlated. A notable exception comes in the lowest model levels (85-91), where perturbations show high mutual correlations. Consideration of Figure~\ref{fig:tends} shows that it is the turbulence and orographic gravity wave drag scheme that is active over these levels. Calculating the correlations separately for land and sea regions shows this enhanced correlation is present for all grid points (not shown). Other regions also show time of day-dependent enhanced correlations. At night time (panel c), the optimal perturbations show substantial correlations between levels 43 and 82. Considering diurnal variations in tendencies indicates that it is the radiation scheme which is active over these levels, with a net cooling tendency at night (see supplementary material figure S8). During the day (panel b), enhanced correlations are found between levels 48 and 61. Figure S8 shows several schemes are active over those levels, but we note that the large scale water processes scheme shows a warming (moistening) tendency confined to that vertical block.

Comparing the vertical regions of enhanced correlation in Figure~\ref{fig:SPPTvert_91_corr} to the distribution of $e_z$ as a function of height in Figure~\ref{fig:SPPTvert_91_disb} reveals that vertical regions with enhanced correlations coincide with vertical regions over which the statistical properties of $e_z$ are reasonably constant in height. It seems justifiable to conclude that a single multiplicative perturbation could be appropriate over those vertical regions. 

\subsection{Coherency of optimal perturbations between different prognostic tendencies} \label{sec:vSPPT}

SPPT assumes that the random error due to the parametrisation process is coherent between the parametrised tendencies for different prognostic variables. Most parametrisation schemes produce tendencies for more than one prognostic variable, the exception being radiation which only impacts $T$. In general the tendencies in different variables are physically related, for example thermodynamic process impact both $T$ and $q$, or convective processes transport momentum, heat, and moisture aloft.
Perturbing the different prognostic tendencies with the same random number is designed to improve the physical consistency of SPPT, representing an instantaneous amplification or reduction of the strength of the parametrised processes. 

To test this assumption, we take the opposite position and consider a generalised `variable SPPT' (vSPPT) in which each prognostic variable, $X$ = $T$, $q$, $U$, $V$, is perturbed with an independent pattern. The optimal perturbation, $e_X$, is fitted independently for each prognostic variable. If the statistical characteristics of the patterns are similar for each prognostic variable, and if there is a high degree of correlation between the $e_X$ fitted to different variable tendencies, then we consider that evidence that a single perturbation should be used for all prognostic tendencies as proposed in SPPT.

The moments and spatio-temporal correlations are computed separately for each $e_X$ following the methodology used in section~\ref{sec:assessSPPT}. These statistics are summarised in Table~\ref{tab:SPPTeX}, showing clear differences to the statistics of $e$ shown in Table~\ref{tab:SPPTe}. The standard deviations of the $e_X$ are higher than the standard deviation of $e$. There is a clear grouping into two pairs: $e_U$ and $e_V$ have a substantially higher standard deviation but smaller spatial decorrelation scales than $e_T$ and $e_q$. The wind perturbations $e_U$ and $e_V$ also have a substantial negative mean, whereas the means for $e_T$ and $e_q$ are close to zero.

\begin{table*}
  \centering
  \begin{tabular}{l | lll | lll | lll | lll}
  \hline
                         &    \multicolumn{3}{l|}{$T$}   &   \multicolumn{3}{l|}{$q$}  & \multicolumn{3}{l|}{$U$}  &  \multicolumn{3}{l}{$V$}      \\
       \hline
     $\mu(e_X)$          &  -0.060  &         &        &  -0.017  &        &       &  -0.37  &        &      &   -0.52 &        &          \\
     $\sigma(e_X)$       &  0.70    &         &        &    0.65  &        &       &   1.7   &        &      &    1.9  &        &          \\
     $\sigma_j$          &  0.66,   &  0.17,  &  0.13  &   0.60,  & 0.22,  & 0.10  & 1.6,    & 0.47,  & 0.18 & 1.8,    & 0.54,  & 0.18     \\
     $L_j$ (km)          &  39,     & 400,    &    --  &  33,     & 430,   &  --   &  28,    & 270    & --   &  26,    & 290,   & --       \\
     $\tau_j$            &  0.6 h,  &  3.5 d, &    --  &  1.2 h,  & 4.3 d, & --    & 1.2 h,  & 3.8 d, & --   & 1.2 h,  & 4.2 d, & --       \\
  \hline
  \end{tabular}
  \caption{As for table~\ref{tab:SPPTe} except treating each prognostic variable ($X$ = $T$, $q$, $U$, $V$) independently.   
  Parameter values are shown for the random fields $j=1, 2, 3$ that comprise the 3-scale pattern similar to that used in the IFS: Standard deviation $\sigma_j$, horizontal correlation length $L_j$, time decorrelation scale $\tau_j$. The spatial and temporal scale of the third pattern cannot be accurately estimated due to the limited size of the domain and length of dataset.}
  \label{tab:SPPTeX}
\end{table*}

Aside from similarities to $e_q$ in terms of mean, standard deviation, and spatial decorrelation scales, the $e_T$ pattern shows certain characteristics not shared by any other $e_X$. The temperature perturbations $e_T$ decorrelate more rapidly in time than $e$ or the other $e_X$. The temperature perturbation $e_T$ also shows a different spatial correlation structure, with a \emph{negative} correlation at a lag of 1.5$^\mathrm{o}$. This is not captured by the AR-1 model (not shown), though at larger spatial lags the AR-1 model is a good representation of the correlation structure of $e_T$. Finally, the standard deviation of $e_T$ is higher over ocean than over land in contrast to $e$ and the other tendencies (see supplementary online material Figure S9).

\begin{figure}
  \centering
  \includegraphics[trim=0.05cm 0cm 0cm 0cm, clip=true, width=0.48\textwidth]{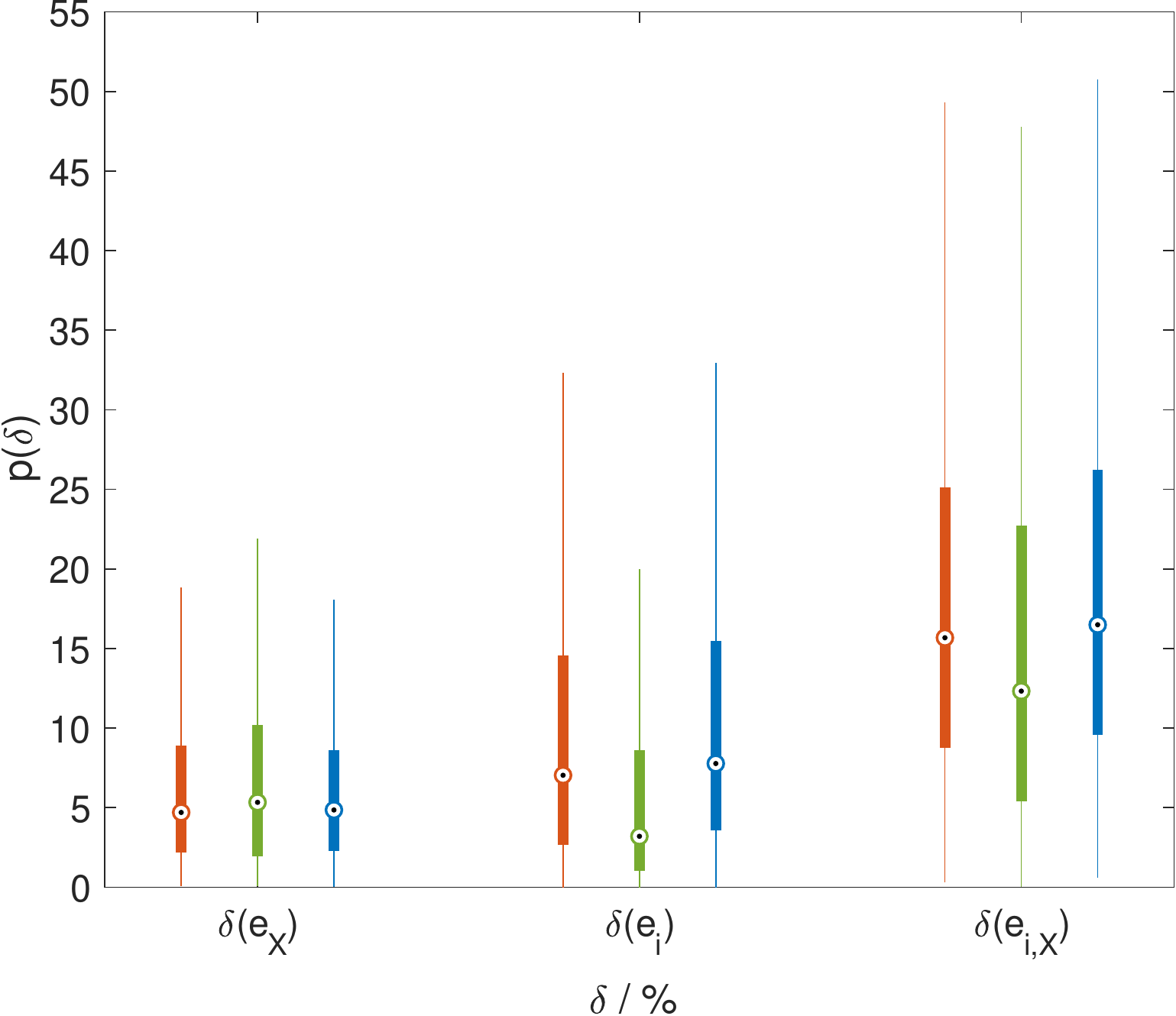}
    \caption{Distribution in the change of the mean square difference between the actual error, $T-D-\Sigma P$ and the modelled error, $e\Sigma P$ for different modifications to SPPT: when independent $e_X$ are used for different variables; when independent $e_i$ are used for different parametrisation schemes; when independent $e_{i,X}$ are used for different parametrisation schemes and variables. $\delta$ is the percentage improvement in the representation of the error on using the modified SPPT approach. The data are shown as box plots, where the median $\delta$s are shown as a circle, and the boxes indicate the 25th and 75th percentiles. Orange: data aggregated from across whole domain. Green: only land points. Blue: only ocean points.}
  \label{fig:SPPTmom_fv}
\end{figure}

Fitting a separate $e_X$ to each prognostic tendency is expected to improve the error characterisation compared to fitting a single pattern because of the additional degrees of freedom in the fitting procedure. To quantify the improvement, the measured error between the SCM and Cascade,
\begin{equation}
 \mathbf{d}_X=\mathbf{T}_X - \mathbf{D}_X -\mathbf{P}_X +\mathbf{b}(\mathbf{P}_X),
\end{equation}
and the modelled errors,
\begin{align}
\mathbf{d}_X^{\mathrm{SPPT}}  &= e \mathbf{P}_X, \nonumber \\
\mathbf{d}_X^{\mathrm{vSPPT}} &= e_X \mathbf{P}_X,  \\
\end{align}
are calculated for each variable, $X$. The mean square difference ($MSD$) between measured and modelled error is calculated for each model and averaged over all variables, as a function of time and spatial position. The percentage improvement,
\begin{equation}
 \delta = 100 \cdot \frac{ MSD^\mathrm{SPPT} - MSD^\mathrm{vSPPT} }{MSD^\mathrm{SPPT}},
\end{equation}
is evaluated. Figure ~\ref{fig:SPPTmom_fv} summarises the distribution of $\delta$ over time and spatial positions using a box and whisker diagram. The median improvement is 5\%, with the whiskers extending to 20\%. The median fractional variance explained by the approach (i.e., the ratio of modelled variance, $\sigma^2(e \mathbf{P}_X)$, to the variance of the measured error, $\sigma^2(\mathbf{d}_X)$ increases by a factor of 2.7 on moving from SPPT to vSPPT (see Supplementary Figure S10).

\begin{figure}
  \centering
  \includegraphics[width=0.48\textwidth]{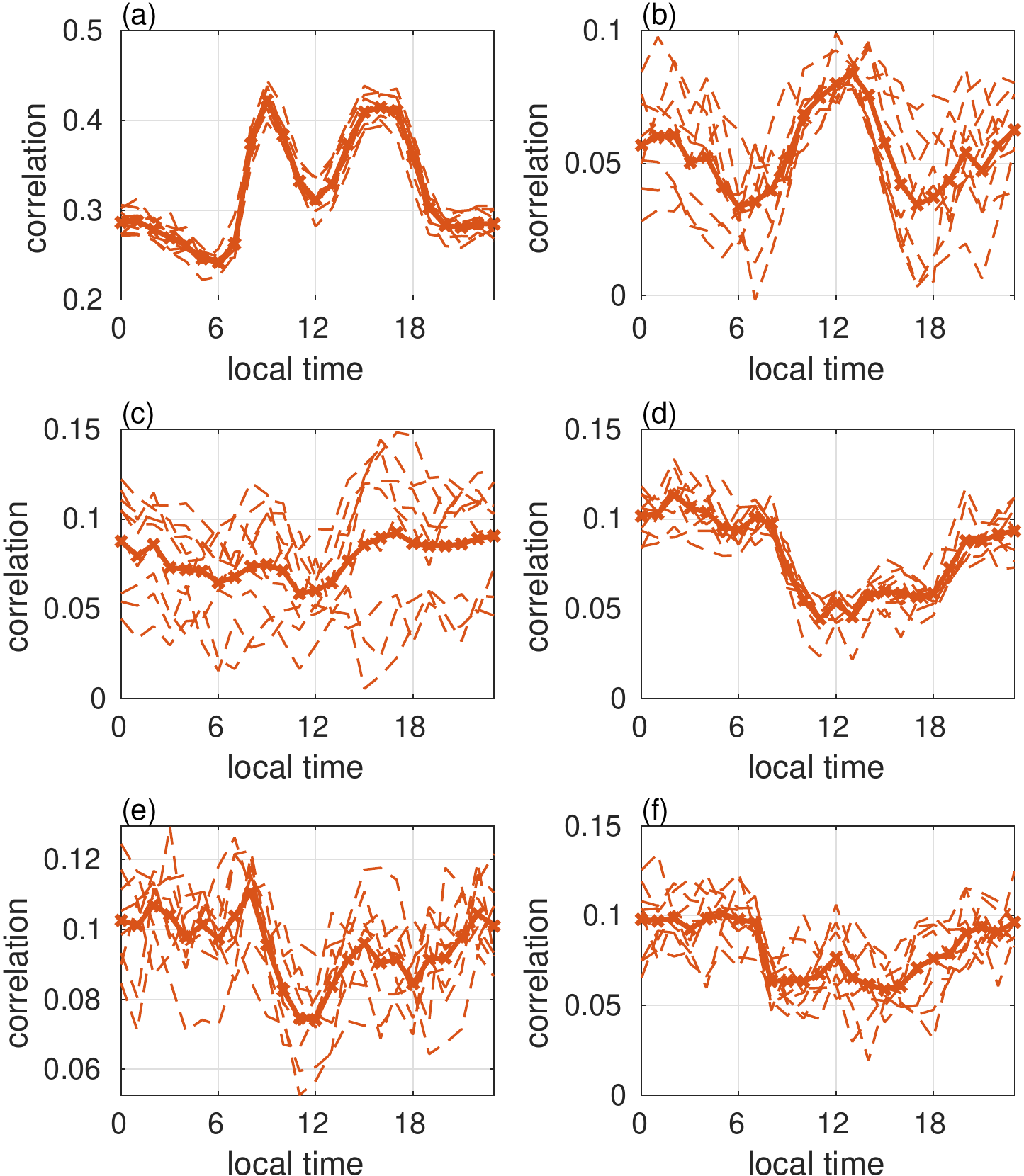}
    \caption{Correlation between $e_X$ fitted to different variable tendencies, as a function of local time of day. (a) $r(e_T, e_q)$; (b) $r(e_T, e_U)$; (c) $r(e_T, e_V)$; (d) $r(e_q, e_U)$; (e) $r(e_q, e_V)$; (f) $r(e_U, e_V)$. The solid line indicates the average statistics over the entire dataset, while the dashed lines indicate the the statistic calculated for each of the nine days separately.}
  \label{fig:SPPT_vari_corr}
\end{figure}

Figure~\ref{fig:SPPT_vari_corr} shows the correlation between $e_X$ fitted for different $X$ as a function of local time of day. Most variable pairs show a modest correlation of order 0.1. However, there are noticeable correlations of 0.3 to 0.45 between $e_T$ and $e_q$, peaking in the mid morning and mid afternoon. This high correlation between $e_T$ and $e_q$ is not unexpected given the physical relationship between temperature and moisture tendencies associated with thermodynamic processes. Consideration of the diurnal cycle in parametrised processes indicates that convective activity peaks in mid morning, while large scale water processes peaks in mid afternoon (supplementary online material Figure S8), explaining the peak in correlation at those times.

Given the high correlations between $e_T$ and $e_q$, and the similarity in standard deviation and spatial correlations between these variable perturbations, perturbing $T$ and $q$ with the same pattern as in SPPT seems physically justified. It is interesting to note that there is not a high correlation between $e_U$ and $e_V$, despite the intimate relationship between $U$ and $V$. It is possible that consideration of wind magnitude and direction, or divergent and rotational flow, would indicate correlated errors in these two variables, not evident when considering $U$ and $V$.

\subsection{Coherency of perturbations between different parametrisation schemes: iSPPT}

The final assumption of SPPT to consider is that the random error is coherent between the different physics parametrisation schemes, such that the balance balance between tendencies from different physics schemes is retained. However this may not accurately reflect variations in model uncertainty \citep{leutbecher2017}. As in section~\ref{sec:vSPPT}, we test this by taking the opposite stance and considering a generalisation to SPPT where each physics parametrisation scheme is perturbed using an independent pattern:
\begin{equation}
\mathbf{T} = \mathbf{D} + \sum_{i=1}^I {(1+e_i)\mathbf{P}_i} - b(\mathbf{P}). \label{eq:iSPPT}
\end{equation}
In this `independent SPPT (iSPPT) approach, proposed by \citet{christensen2017b}, the spatio-temporal characteristics of the independent patterns $e_i$ can be specified by the user, allowing for the representation of different model error characteristics associated with each physical parametrisation scheme. We can use the coarse-graining framework to assess whether the statistical characteristics of the $e_i$ are indeed different, or if using a single $e$ as in SPPT is sufficient to represent model uncertainty in the IFS.

The $e_i$ are estimated by solving the over-constrained vector equation:
\begin{equation}
	\mathbf{T} - \mathbf{D} -\mathbf{P} +\mathbf{b}(\mathbf{P}) = \underline{\underline{\mathbf{P}}} \mathbf{e} \label{eq:SPPTei}
\end{equation}
at every spatial location and time step, where the matrix $\underline{\underline{\mathbf{P}}}$ consists of $I$ columns each containing parametrised tendency $i$ as a function of height. The $I \, \mathrm{x}\,1$ vector $\mathbf{e}$ contains the $e_i$ optimal perturbations. It was found that fitting an independent perturbation for non-orographic gravity wave drag (NOGW) led to instabilities in the fitting procedure, because that scheme produces tendencies that are three order of magnitude smaller than the other parametrisation schemes (Figure~\ref{fig:tends}). The NOGW tendencies were therefore excluded from $\underline{\underline{\mathbf{P}}}$ and instead moved to the left hand side of equation~\ref{eq:SPPTei}, and $\mathbf{e}$ calculated for the remaining four schemes.

\begin{figure*}
  \centering
  \includegraphics[trim=0.5cm 0.5cm 0.5cm 0.0cm, clip=true, width=0.98\textwidth]{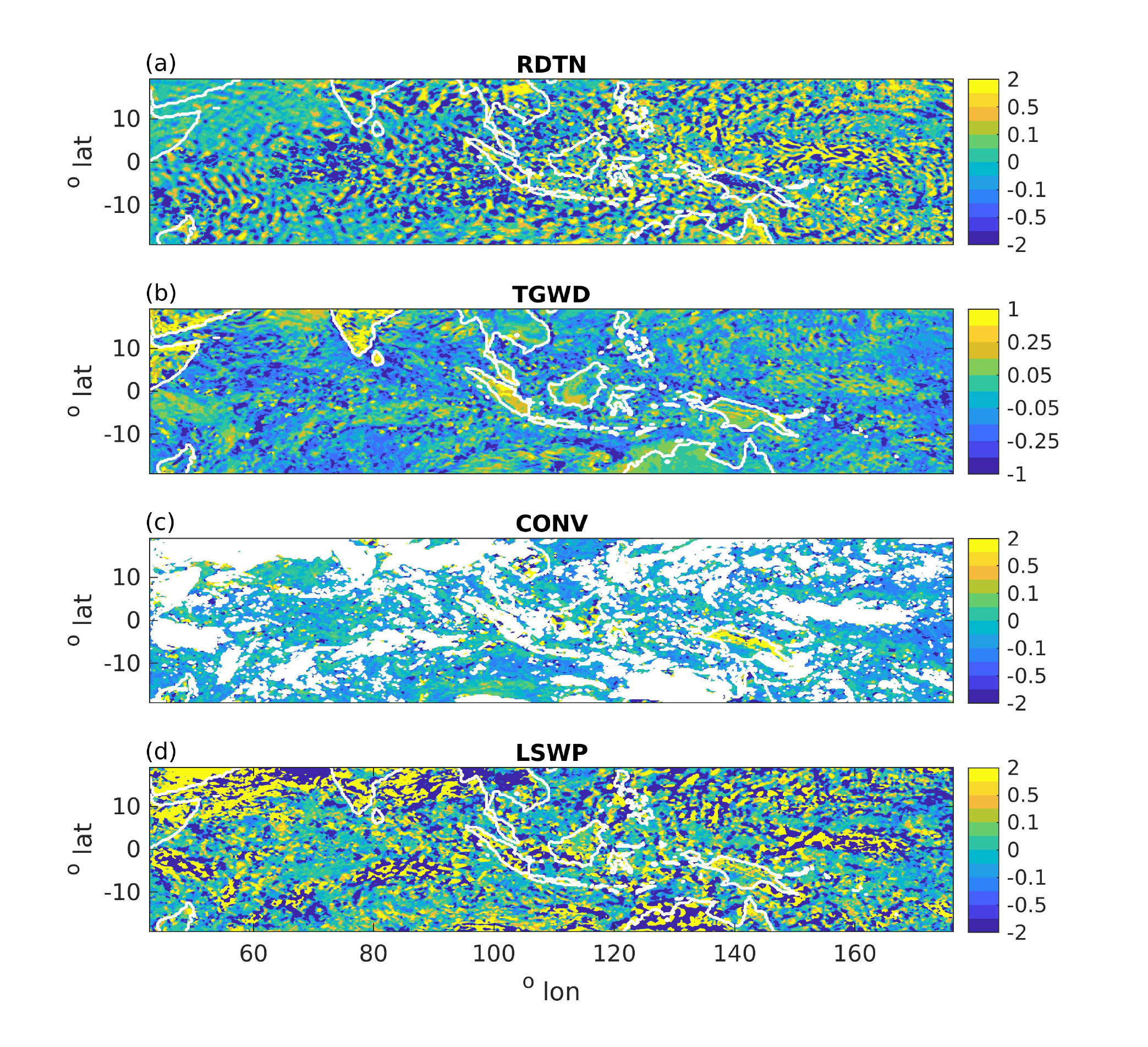}
    \caption{A snapshot of the optimal multiplicative perturbations, $e_i$, as defined in the iSPPT framework. The SCM forecasts were initialised at 00UTC, 7 April 2009.}
  \label{fig:iSPPTei}
\end{figure*}

Figure~\ref{fig:iSPPTei} shows a simultaneous snapshot of the optimal $e_i$ for each of the physical parametrisation schemes considered: radiation (RDTN), turbulence and gravity wave drag (TGWD), convection (CONV), and large scale water processes (LSWP). A number of interesting results are apparent. Firstly, different parametrisation schemes show different error characteristics: for example, the optimal perturbation to the radiation scheme appears to have significantly smaller scales than the other schemes. Secondly, in regions where the convection scheme did not trigger (white in panel three), the perturbation in both the radiation and large scale water processes schemes are of larger magnitude than in regions where convection did trigger. This could indicate that the convection parametrisation scheme has uncertainties that are not well represented by (independent) SPPT, for example, errors in triggering of convective events. The characteristics of the perturbation to the turbulence and gravity wave drag scheme are remarkably similar to the characteristics of the $e$ fitted to the total net tendency.

Table~\ref{tab:SPPTei} shows the moments and spatio-temporal correlations calculated for each $e_i$. The perturbations fitted to the TGWD scheme show similar statistics to the SPPT $e$ perturbation, including similar standard deviation, spatial and temporal correlations, and relative magnitudes of the three patterns. The $e_i$ fitted to other parametrisation schemes show substantially different statistics. The radiation perturbations have a larger standard deviation than $e$. The first pattern explains substantially more variance, explaining the shorter correlation scales observed for $e_\mathrm{RDTN}$ in Figure~\ref{fig:iSPPTei}. Inspection of the full spatial decorrelation structure shows AR-2 behaviour similar to that observed for $e_T$: it is likely that the AR-2 behaviour in $e_T$ can be traced to uncertainty in the radiation scheme. The perturbations $e_\mathrm{CONV}$ and $e_\mathrm{LSWP}$ also show larger standard deviations and a larger weighting on the first pattern than $e$, though they also show larger temporal correlation scales. The standard deviation of $e_\mathrm{LSWP}$ is very large. This can be traced to regions where the convection scheme has not triggered.

\begin{figure}
  \centering
  \includegraphics[width=0.48\textwidth]{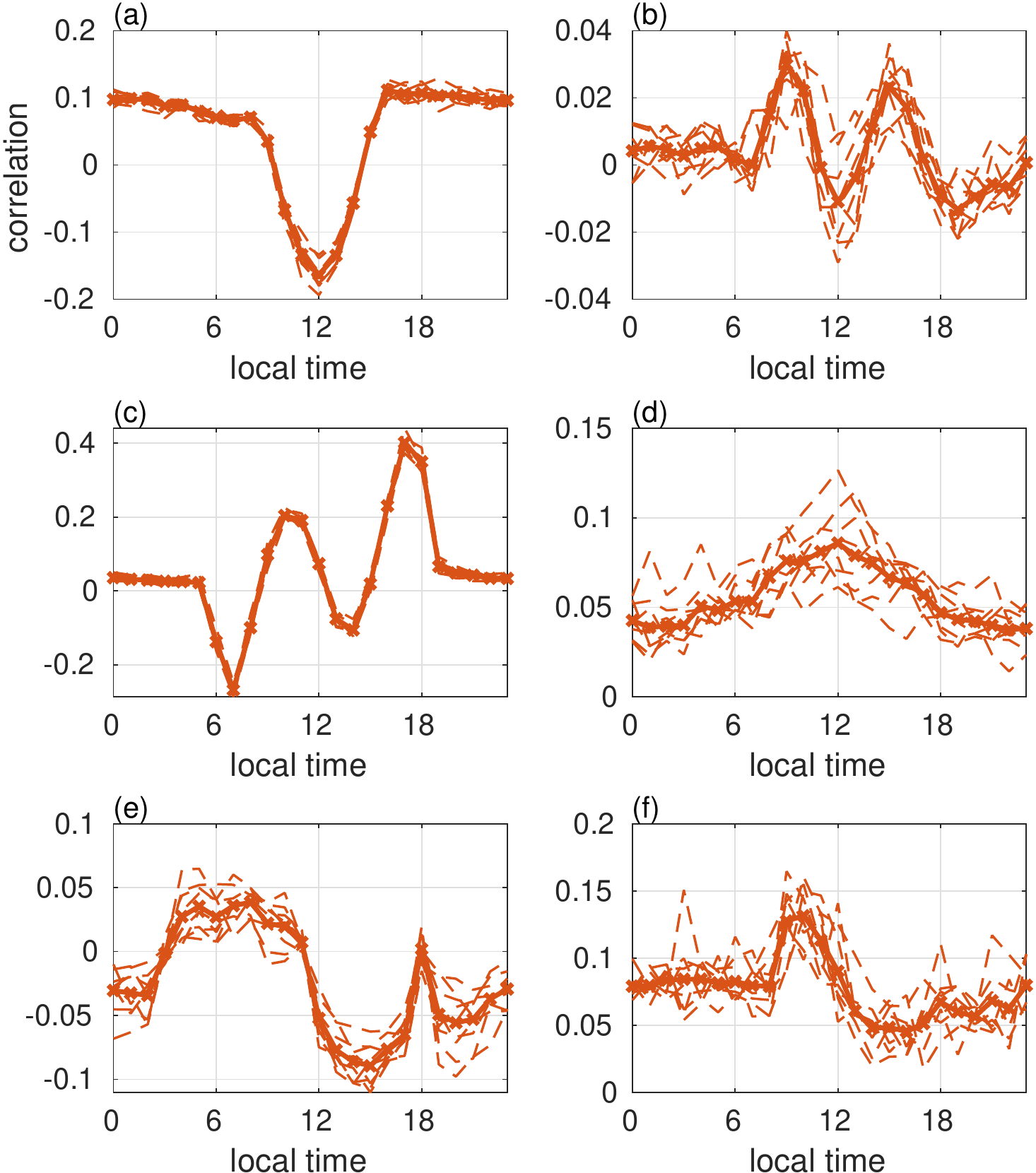}
    \caption{Correlation between $e_i$ fitted to different parametrisation schemes, as a function of local time of day, as defined in the iSPPT framework. (a) $r(e_{\mathrm{RDTN}}, e_{\mathrm{TGWD}})$; (b) $r(e_{\mathrm{RDTN}}, e_{\mathrm{CONV}})$; (c) $r(e_{\mathrm{RDTN}}, e_{\mathrm{LSWP}})$; (d) $r(e_{\mathrm{TGWD}}, e_{\mathrm{CONV}})$; (e) $r(e_{\mathrm{TGWD}}, e_{\mathrm{LSWP}})$; (f) $r(e_{\mathrm{CONV}}, e_{\mathrm{LSWP}})$. The solid line indicates the average statistics over the entire dataset, while the dashed lines indicate the the statistic calculated for each of the nine days separately.}
  \label{fig:iSPPTei_corr}
\end{figure}

\begin{table*}
  \centering
  \begin{tabular}{l | lll | lll | lll | lll}
  \hline
                     &    \multicolumn{3}{l|}{RDTN}   &   \multicolumn{3}{l|}{TGWD}  & \multicolumn{3}{l|}{CONV} &  \multicolumn{3}{l}{LSWP (in CONV regions)}  \\
  \hline
     $\mu(e_X)$      & -0.023   &           &         & -0.043  &          &         &  -0.16   &        &          &  \multicolumn{3}{l}{-0.34 (0.18)}              \\
     $\sigma(e_X)$   & 1.4      &           &         & 0.66    &          &         &  4.0     &        &          &  \multicolumn{3}{l}{14 (2.7)}                  \\
     $\sigma_j$      &  1.4,    & 0.32,     & 0.24    & 0.59,   & 0.28,    & 0.11    & 3.9,     & 0.87,  & 0.15,    & 14(2.6),  & 3.2(0.59), & 1.8(0.34)         \\
     $L_j$ (km)      &  38,     &  570,     &  --     & 27,     & 330,     & --      & 16,      & 240,   & --       &  33,      & 370,        & --                 \\
     $\tau_j$        &  0.79h,  & 4.2d      & --      & 1.4h,   & 5.0d     & --      &  0.72h,  & 5.6d,  & --       &  0.86h,   & 6.5d       &  --                \\
  \hline
  \end{tabular}
  \caption{As for table~\ref{tab:SPPTeX} except treating each parametrisation scheme (RDTN, TGWD, CONV, LSWP) independently. The bracketed numbers in the LSWP column indicate the pattern parameters if analysis is only carried out in regions where the convection parametrisation has triggered.
  Parameter values are shown for the random fields $j=1, 2, 3$ that comprise the 3-scale pattern similar to that used in the IFS: Standard deviation $\sigma_j$, horizontal correlation length $L_j$, time decorrelation scale $\tau_j$. The spatial and temporal scale of the third pattern cannot be accurately estimated due to the limited size of the domain and length of dataset.}
  \label{tab:SPPTei}
\end{table*}

The correlation between the perturbations fitted to different schemes as a function of local time of day is shown in Figure~\ref{fig:iSPPTei_corr}. Correlation are generally small, but show a clear diurnal structure. The correlations between the perturbations fitted to different schemes are due to physical relationships between the schemes. As an example, consider the temporally varying relationship between $e_{\mathrm{RDTN}}$ and $e_{\mathrm{LSWP}}$ as shown in Figure~\ref{fig:iSPPTei_corr}(c). Between 7pm and 5am the correlation is zero. From 6am we see an initial negative correlation, before two periods (9-12am and 4-6pm) with a substantial positive correlation of up to 0.4.

The formation of stratiform cloud as parametrised in the IFS is directly dependent on previous heating tendencies from radiation and other diabetic processes. Radiative heating increases the saturation specific humidity, either reducing the condensation rate or leading to evaporation of the cloud. If the radiative tendencies are positive, and the random error in radiative tendencies is also positive ($e_{\mathrm{RDTN}} > 0$), this indicates radiation in Cascade warms the column more than in the SCM. The sign of the error in the LSWP tendencies will depend on whether the SCM LSWP scheme indicates formation or evaporation of cloud in that region. If $\mathrm{P}_{T, \mathrm{LSWP} }>0$, condensation is occurring, and a higher radiative warming would have reduced this tendency. It is expected that $e_{\mathrm{LSWP}} < 0$. Conversely if $\mathrm{P}_{T, \mathrm{LSWP} }<0$, evaporation is occurring, and a higher radiative warming would enhance this tendency: $e_{\mathrm{LSWP}} > 0$. Similar arguments can be made for negative radiative tendencies. We conclude that the correlation between error in radiative tendencies and error in cloud tendencies is dependent on the sign of the tendencies: if the tendencies have the same sign, we expect the errors to be negatively correlated, and vice versa.

\begin{figure*}
  \centering
  \includegraphics[width=0.78\textwidth]{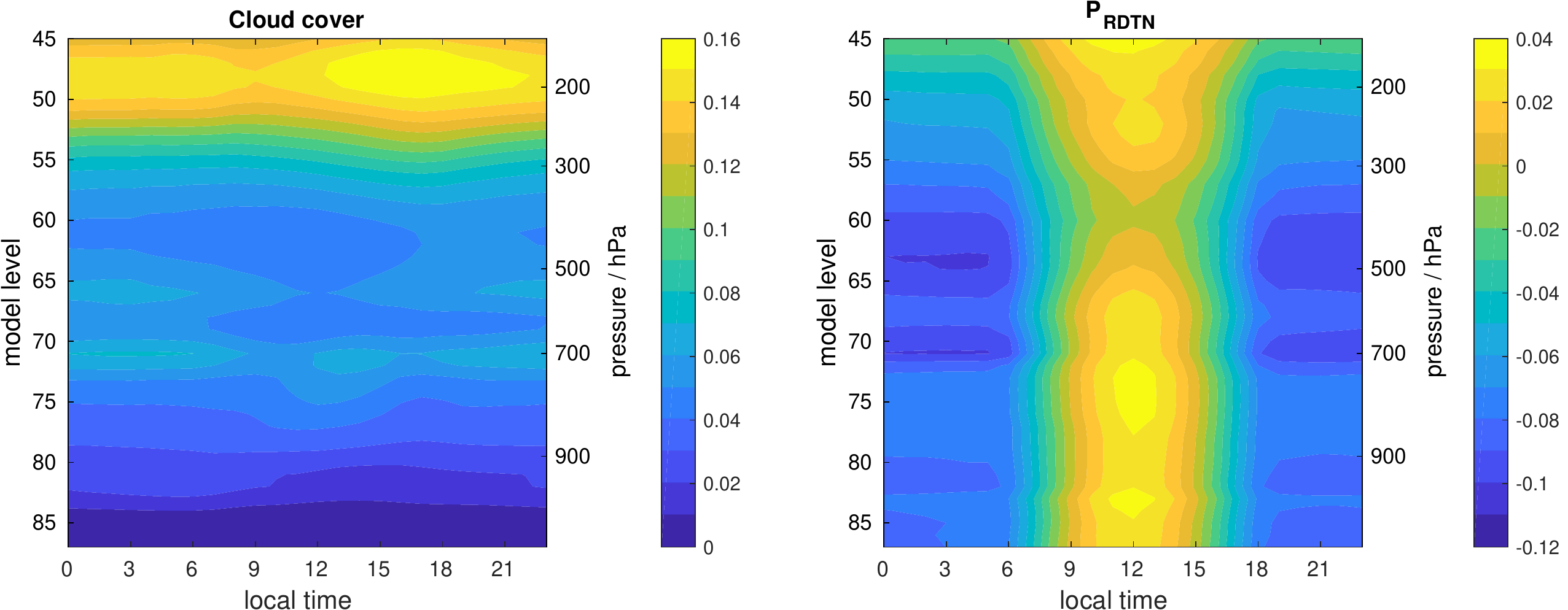}
    \caption{Mean cloud cover (fraction) and mean radiative tendencies (K h$^\mathrm{-1}$) as modelled by the SCM, as a function of model level and local time of day. The average is computed over all start times and all grid points.  The right hand vertical axis shows characteristic pressure levels for reference.}
  \label{fig:rdtt_cloud}
\end{figure*}

Figure~\ref{fig:rdtt_cloud} indicates that radiative tendencies are only positive during the middle of the day, between 9 am and 3 pm. Between 6 am and 10 am -- 12 noon (dependent on height), Figure~\ref{fig:rdtt_cloud} shows that the SCM indicates decaying cloud cover, associated with a negative temperature tendency, $\mathrm{P}_{T, \mathrm{LSWP} }<0$. Before 9am, the radiative and cloud tendencies have the same sign, whereas from 9am onwards, the signs are opposite, hence the change in correlation between $e_{\mathrm{RDTN}}$ and $e_{\mathrm{LSWP}}$ at this time. From 10 am to 6 pm, the SCM predicts an increase in high-level cloud cover associated with positive temperature tendencies. Before 3 pm, the cloud and radiative tendencies again have the same sign, such that errors are anti-correlated. After this, the radiation tendency flips sign, and the positive correlation in $e$ returns.

As for vSPPT, fitting a separate $e_i$ to each parametrisation scheme is likely to improve the error characterisation because of the additional degrees of freedom in the fitting procedure. Figure ~\ref{fig:SPPTmom_fv} summarises the improvement in fit due to iSPPT over SPPT in terms of the distribution of $\delta$. The median improvement is 7\%, with the whiskers extending to over 30\%. If the vSPPT and iSPPT approaches are combined (ivSPPT), and a separate $e_{i,X}$ fitted for each parametrisation scheme for each variable, the median improvement increases to 15\%, with tails extending to 50\%. Similarly, the median fractional variance explained by the approach increases by a factor of 3.4 (5.0) on moving from SPPT to iSPPT (ivSPPT) (see Supplementary Figure S10). For both iSPPT and ivSPPT, the improvement is larger over ocean points than over land points. The remaining residual between measured and modelled error cannot easily be represented through a multiplicative approach.


\section{Discussion} \label{sec:discussion}


Coarse-graining is an attractive technique for characterising model error. A high resolution simulation, which resolves the processes of interest, is taken as a benchmark. The simulation is coarsened to the resolution of the forecast model, before its evolution is compared to forecasts made with the low resolution model. The difference between the coarse-grained high resolution simulation and the forecast is considered the `error' in the forecast. There are many benefits of using a high-resolution simulation for coarse-graining studies, such as excellent spatio-temporal coverage, and the availability of all model fields including those poorly constrained by observations (e.g. vertical velocity) \citep{christensen2018b}. However, it is important to note that the high-resolution simulation is only a proxy for the truth. The high-resolution simulation must be thoroughly validated against observational data before coarse-graining. This ensures any biases in the high-resolution simulation are understood, to avoid conflation with the forecast `errors' assessed in the coarse-graining analysis. 

This paper has proposed a new technique within the coarse-graining framework for assessing systematic and random model error. Instead of requiring the user to simultaneously produce a pair of low- and high-resolution simulations, it makes use of existing high-resolution simulations. This sets a relatively low bar for carrying out such coarse-graining studies, and makes use of the wealth of high-resolution simulations available \citep{heinze2017,satoh2014,satoh2017,schalkwijk2015}. The coarse-grained dataset is coupled to a forecast model through the use of a single column model (SCM). The initial conditions and forcing fields (including the dynamics tendencies) required by the SCM are provided by the coarse-grained dataset. The SCM is run independently in each coarse-grained grid box and provides the physics tendencies as a function of model level. This allows for a 3D characterisation of error in the parametrised physics tendencies.

The technique can be used to study the development of systematic biases between the high resolution simulation and the low resolution forecast model, developing over just a few time steps. Evaluating bias development over short time periods ensures that any biases identified are unlikely to be due to a remote source. This contamination is a problem when using standard validation techniques, e.g. the evaluation of parametrisation schemes in long climate simulations, where errors from remote sources can compensate or be conflated with errors arising in the region of interest. The presence of compensating errors is particularly problematic for the parametrisation development process, where it can be hard to assess the benefits of a new scheme if that scheme no longer compensates for other errors in the model. Localisation of errors in space and time enables identification of the sources of those errors.

In this study, the focus was on using the coarse-graining approach to characterise and understand random errors in a low resolution forecast model: the ECMWF Integrated Forecasting System (IFS) at resolution $\mathrm{T_L}$639. To illustrate the approach, the widely used Stochastically Perturbed Parametrisation Tendencies (SPPT) scheme was taken as an example of a stochastic parametrisation. Despite its widespread use and beneficial impacts in weather and seasonal forecasts, there is limited evidence supporting the theoretical foundations of the scheme. Through characterising the statistics of random model error in low resolution forecasts, this study seeks to indicate whether the theoretical foundations of SPPT are sound.

We find evidence that uncertainty in the parametrised tendencies increases with the magnitude of the parametrised tendency. Multiplicative noise is therefore a good first-order model for uncertainty in the tendencies, providing support for the use of SPPT. To inform the properties of the stochastic perturbation to be used in SPPT, we calculate an optimal multiplicative perturbation as a function of space and time, and assess its statistical properties. The standard deviation of the optimal perturbation is approximately 30\% smaller than that used operationally at ECMWF. However, the optimal perturbation also has positive skewness and positive excess kurtosis. This results in a distribution with fatter tails than the normal distribution used at ECMWF, and would increase the frequency of large-magnitude perturbations. The positive skewness also indicates a reduced likelihood of large negative perturbations which change the sign of the parametrised tendency. At ECMWF, the SPPT perturbations are truncated at plus or minus one\footnote{Note that this truncation reduces the actual standard deviation of the perturbation by around 6\% from that stated.} to ensure SPPT does not invert the sign of the tendency. Using a skewed distribution would reduce the need for this truncation. Overall, it seems the estimated characteristics of SPPT are not very different to those used operationally. It is reassuring to find that the `top down' approach of tuning the SPPT scheme to produce reliable forecasts is able to find similar parameter values to those found in this `bottom up' coarse-graining approach. 

Importantly, the optimal multiplicative perturbation was found to be correlated in space and time. Spatio-temporally correlated noise has long been recognised as necessary for a skilful stochastic parametrisation scheme \citep{buizza1999}, and while this can be motivated by theoretical considerations, no coarse-graining studies have presented evidence that this is physically justified. This coarse-graining study provides that evidence. Each SCM simulation is produced independently from its space-time neighbours. Any spatio-temporal correlations in the optimal perturbation must be due to correlated errors in SCM behaviour under different meteorological or boundary conditions. For example, wind shear may introduce convective organisation which is not well represented by the parametrised model, leading to correlated errors in convective tendencies. The use of spatio-temporally correlated noise in stochastic parametrisations allows for the representation of such errors. The decorrelation scales were estimated in space and time and compared to those used in SPPT. The estimated decorrelation in space is more rapid than used in SPPT, though in time the estimated $e$ showed higher correlations at long lags than in the operational scheme. There is the potential for these estimated correlation scales to be used in SPPT. However, it is not clear to what extent the statistics would change for regions other than the Tropical Pacific, and for other time periods. Furthermore, the study is limited by the spatio-temporal domain of the high-resolution simulation, which restricts the ability to estimate correlations on the largest space and time scales. It is possible that the longest space- and time-scales used in operational SPPT actually represent errors due to the coupled ocean-atmosphere system. These errors cannot be assessed in this framework. In any case, such errors would be better represented by including stochasticity into the ocean model.

This study also indicates several limitations of SPPT. The optimum perturbation shows substantial variation in its properties between land and sea points in terms of moments of the perturbation and correlation characteristics. In addition, over land the perturbation shows a marked diurnal cycle, with a substantially higher standard deviation at night than during the day (Figure~\ref{fig:SPPTmom}). This reveals a limitation of the SPPT approach. At night over land, the parametrised physics tendencies are smaller in magnitude than during the day. However, in this case, the uncertainty in the tendencies does not reduce proportionally, such that the multiplicative perturbation must be amplified to represent uncertainty in the small-magnitude tendencies at night. It is possible that an additional, state independent uncertainty representation could be appropriate here. Other approaches could also be used to improve the representation of the diurnal cycle in uncertainty. For example, \citet{lock2019} highlight a recent development to SPPT which accounts for low uncertainty in clear sky radiative tendencies. This leads to differing representations of uncertainty as a function of time of day, due to diurnal variations in the radiative tendencies.

We see further evidence that uncertainty in the IFS is not perfectly multiplicative in Figure~\ref{fig:hist_manylev}. Between levels 66 and 52 (555--240 hPa), uncertainty in positive temperature tendencies is markedly non-linear, increasing at a slower rate than linear. By considering the levels at which different parametrisation schemes are active, these features can be attributes to a particular physical process: the nonlinear relationship at these levels is due to uncertainty in the convection parametrisation. A non-linear relationship between uncertainty in convection and the magnitude of convective tendencies was also highlighted by \citet{shutts2014}. Several alternative approaches have been proposed to represent uncertainty in convection parametrisation. Of particular note is the Plant-Craig scheme \citep{plant2008}. The underlying theory was proposed with tropical convection in mind \citep{craig2006}, but the generality of the theory for convection over land has recently been demonstrated \citep{rasp2018}. The Plant-Craig approach represents the uncertainty in the convective mass flux as proportional to the square root of its mean. Many convection parametrisation schemes begin by estimating the mass-flux as a function of stability, convectively available potential energy or moisture budgets, before this is used as an input to an entraining parcel model. While in this study, the uncertainty in the \emph{output} of the convection parametrisation is considered, as opposed to the uncertainty in what is effectively an \emph{input}, there seems to be a consistency between our results and the Plant-Craig approach.

In section~\ref{sec:beyondSPPT}, the coarse-graining analysis was used to assess three further assumptions made in SPPT. Firstly, we consider the assumption that the SPPT perturbation is constant in the vertical, such that the whole vector tendency is scaled up or down. To assess the justification for this, a separate multiplicative perturbation was fitted at each vertical level, and the correlation calculated between perturbations fitted at different levels. In general, correlations were weak, with the strongest correlations found between levels affected by the same parametrisation schemes. This was particularly evident for the boundary layer scheme, and for night time radiative tendencies. Over those levels, the statistics of the optimal perturbation were also found to be approximately constant. The use of a constant perturbation in height ensures consistency is maintained for schemes which represent transport processes: the whole tendency is scaled up or down to ensure conservation of mass and tracers. It seems that using a constant perturbation for each parametrisation scheme would be sufficient, instead of for the whole vertical column. 

The second assumption considered is the use of a single perturbation for all prognostic variable tendencies. As for vertical coherence, this assumption was tested by relaxing the assumption. The statistics of the optimal perturbation fitted to each variable tendency were then considered. The optimal perturbation fitted to $q$ and $T$ was found to be correlated throughout the day. For moist processes, changes in $q$ are associated with a change in $T$, explaining the correlation in errors for these tendencies. We would not expect this correlation to be perfect as dry processes can also change $T$. We would also expect the zonal and meridional wind tendencies to be related. However, the correlation between errors in these tendencies is small. It is possible that expressing the wind as stream function and velocity potential would be informative. The statistical properties of the optimal perturbation fitted to $U$ and $V$ tendencies are similar, and the perturbations fitted to $T$ and $q$ are also similar. This motivates the development of stochastic parametrisations separately for thermodynamic and dynamic processes.

The final SPPT assumption considered in this study is that the error is coherent between all physical parametrisation schemes. This was assessed by fitting a separate perturbation to each parametrisation scheme. The correlation between different schemes was generally weak, though the radiation perturbation showed robust diurnally-varying correlations with both the cloud and boundary layer schemes, due to physical relationships between these processes. The perturbations fitted to separate schemes were found to have markedly different statistical characteristics, including magnitude of perturbation and correlation scales. Together with the generally weak correlations between perturbations fitted to different schemes, and the assessment of vertical coherence of perturbations, we find support for the `independent SPPT' (iSPPT) approach proposed by \citet{christensen2017b}. It is known that the iSPPT approach improves forecast reliability in the IFS, and that it has its largest beneficial impact in the tropics \citep{christensen2017b}. This study considers a tropical domain, so it is again reassuring that the `bottom up' approach of estimating instantaneous error statistics reaches the same conclusions as the `top down' approach of assessing medium-range forecast reliability. However it is possible that if an extra-tropical domain were used for coarse-graining analysis, the SPPT approach would appear more favourable compared to iSPPT.  

Finally, even allowing for independent perturbations to each scheme and to each variable, it is not possible to fully account for uncertainty in the parametrised tendencies using a multiplicative approach. Allowing for these generalisations results in an improvement of the fit between modelled and measured error by up to 50\% over SPPT, but the average improvement is only 15\%. This indicates that there are model errors in the IFS which cannot be represented using multiplicative noise. This motivates the continued development of new stochastic parametrisations to better characterise model error.

While the Cascade simulation has been thoroughly validated against observations, it is possible that the results presented here are sensitive to errors in the truth simulation, or to other details of the truth model. While assessing this sensitivity is outside the scope of this study, future work will evaluate the sensitivity of the results to the truth model. Other details of the experimentation will also be considered, including a comparison between different forecast models, domains, and meteorological conditions.

\section{Conclusions and Recommendations} \label{sec:concs}

We conclude by suggesting some recommendations for stochastic parametrisation based on this study, in order of priority:
\begin{enumerate}
  \item There is evidence that multiplicative noise is a reasonable first-order representation of uncertainty in the IFS parametrised tendencies, providing some support for the use of SPPT. However, the evidence also suggests that the convection scheme in particular could benefit from an alternative approach, such that the uncertainty in the convection tendencies increase at a rate slower than linear.
  \item The spatio-temporal correlations used in stochastic parametrisation schemes such as SPPT have a physical basis, and are not only necessary for pragmatic reasons.
  \item The standard deviation of perturbations used in SPPT should be reduced, but the random perturbations should also be drawn from a skewed distribution. This will reduce the need for truncating the distribution to avoid negative perturbations. 
  \item The iSPPT approach \citep{christensen2017b} seems to account for many of the results shown, including. 
  \begin{enumerate}
    \item The correlation between perturbations at different vertical levels is limited to within parametrisations.
    \item A low correlation is found between perturbations fitted to different parametrisation schemes.
    \item Perturbations fitted to different schemes show very different noise characteristics. These different  model error characteristics can be specified within the iSPPT approach.
    \item The correlations between perturbations applied to different variables are due to the physical relationship between these variables, as represented in the parametrisation schemes.
  \end{enumerate}
This approach also enables multiplicative noise to be easily replaced by alternative model uncertainty representations for specific schemes.
  \item There is some evidence that uncertainty in the thermodynamic ($T$, $q$) and dynamic ($U$, $V$) tendencies should be treated differently, as they exhibit different error characteristics. The optimal method for achieving this is left for future research.
\end{enumerate}

\section{Acknowledgements} 

The research of H.M.C. was supported by European Research Council grant number 291406 and Natural Environment Research Council grant number NE/P018238/1. The author would like to express her particular thanks to Andrew Dawson (ECMWF) for his extensive input into writing software used in this work. The author would also like to thank Chris Holloway (University of Reading) for providing the Cascade data used here, and advising on its use. Thanks also to Tim Palmer (University of Oxford), Judith Berner (NCAR), Martin Leutbecher and Sarah-Jane Lock (both ECMWF) for helpful advice and input into this work. Thanks to Filip Vana (ECMWF) for support with using the IFS SCM. The author is grateful to the ECMWF OpenIFS project (https://www.ecmwf.int/en/research/projects/openifs) for providing access to the IFS SCM. The coarse-grained data used and produced in this study are archived at the Centre for Environmental Data Analysis (http://catalogue.ceda.ac.uk/ uuid/bf4fb57ac7f9461db27dab77c8c97cf2).

\appendix


\section{Estimating the spatial and temporal correlation coefficients}


The optimal perturbation, $e(\phi,\lambda, t)$ is modelled as a sum over $N$ AR1 processes, separately for each spatial dimension (longitude, $\phi$, and latitude, $\lambda$) and in time ($t$). For illustration, consider the time decomposition:
\begin{align}
  e(t) &= \sum_{i=1}^N X_i(t), \\
  X_i(t) &= \phi_i X_i(t-1) + \sigma_i (1-\phi_i^2)^{\frac{1}{2}}\xi
\end{align}
where $\phi_i$ and $\sigma_i$ are the lag-1 autocorrelation and standard deviation of the $i\mathrm{th}$ scale respectively, and $\xi$ is white noise, $\xi \sim \mathcal{N}(0,\,1)$. The $X_i$ are ordered such that the first scale decorrelates the fastest, and the $N\mathrm{th}$ scale decorrelates the slowest. Since the $X_i$ are uncorrelated, the variance and autocorrelation of $e$ can be written:
\begin{align}
\sigma_e^2 &= \sum_{i=1}^N \sigma_i^2 \\
\rho_e     &= \frac{\sum_{i=1}^N \sigma_i^2 \phi_i^\tau}{\sum_{i=1}^N \sigma_i^2}
\end{align}

To select the optimal number of scales, the log of the autocorrelation of $e$ is plotted, revealing a number, $N$, of straight-line sections. For large $\tau$, we assume $\phi_i << \phi_N, i \neq N$, and approximate the autocorrelation as
\begin{align}
\rho_e     &= \frac{ \sigma_N^2 \phi_N^\tau}{\sum_{i=1}^N \sigma_i^2},
\end{align}
In this way, the variance ratio, $\frac{ \sigma_N^2 }{\sum_{i=1}^N \sigma_i^2}$, and autocorrelation, $\phi_N$, of the largest scale can be estimated from the graph of the log of the autocorrelation function at large $\tau$. The modelled $\rho_N$ is subtracted from $\rho_e$, and the method repeated for each of the next slowest scales in turn.

\bibliography{journals,BIBLIOGRAPHY}
\bibliographystyle{agsm}



%


\end{document}